\documentclass[12pt,preprint]{aastex}
\newcommand{\e}{SN~2000E}

\shorttitle{Supernova 2000E}
\shortauthors{Valentini et al.}

\begin{document}
\title{Optical and near-infrared photometry of the Type Ia Supernova 2000E in 
NGC~6951}
\author{G. Valentini\altaffilmark{1}, E. Di Carlo\altaffilmark{1}, 
F. Massi\altaffilmark{2}, M. Dolci\altaffilmark{1}, 
A. A. Arkharov\altaffilmark{4}, V. M. Larionov\altaffilmark{4,8,9},\\ 
A. Pastorello\altaffilmark{5}, A. Di Paola\altaffilmark{3}, 
S. Benetti\altaffilmark{5}, E. Cappellaro\altaffilmark{6}, 
M. Turatto\altaffilmark{5}, F. Pedichini\altaffilmark{3},\\ 
F. D'Alessio\altaffilmark{3}, A. Caratti o Garatti\altaffilmark{3,10}, 
G. Li Causi\altaffilmark{3}, R. Speziali\altaffilmark{3},\\
I. J. Danziger\altaffilmark{7}, A. Tornamb\'e\altaffilmark{1}}
\email{dicarlo@te.astro.it}
\altaffiltext{1}{INAF-Osservatorio Astronomico di Collurania-Teramo, Via M. 
Maggini, I-64100 Teramo, Italy}
\altaffiltext{2}{INAF-Osservatorio Astrofisico di Arcetri, Largo E. Fermi 5, 
I-50125 Firenze, Italy}
\altaffiltext{3}{INAF-Osservatorio Astronomico di Roma, Via Frascati 33, 
I-00040 Monteporzio Catone (Roma), Italy} 
\altaffiltext{4}{Central Astronomical Observatory at Pulkovo, Pulkovskoe 
shosse 65, 196140 Saint-Petersburg, Russia}
\altaffiltext{5}{INAF-Osservatorio Astronomico di Padova, Vicolo 
dell'Osservatorio 5, I-35122 Padova, Italy}
\altaffiltext{6}{INAF-Osservatorio Astronomico di Capodimonte, via Moiariello 
16, I-80191 Napoli, Italy}
\altaffiltext{7}{INAF-Osservatorio Astronomico di Trieste, Via G.B. Tiepolo 
11, I-34131 Trieste, Italy}  
\altaffiltext{8}{Astronomical Institute of St. Petersburg University, Russia}
\altaffiltext{9}{Isaac Newton Institute of Chile, St. Petersburg branch}
\altaffiltext{10}{Universit\'a degli Studi di Roma ``Tor Vergata'' - 
Dipartimento di Fisica, via Della Ricerca Scientifica 1, I-00133 Roma, Italy}

\begin{abstract}
We present optical ($UBVRI$) and near-infrared ($JHK$) 
photometry, along with optical spectra, of the Type Ia supernova SN~2000E 
in the spiral galaxy NGC~6951. It was discovered by the staff of the
Teramo Observatory during the monitoring of the SN~1999el.
The observations span a time interval of 234 days in the optical
 and 134 days in the near-infrared (starting $\sim 16$ days and 
$\sim 7$ days before maximum $B$ light, respectively). Optical 
spectra are available from 6 days before maximum $B$ light to 122 days after 
it.
The photometric behavior of SN~2000E is remarkably similar to that of 
SN~1991T and SN~1992bc: it exhibits a $\Delta m_{15}(B) = 0.94$, thus being 
classifiable as a slow-declining Type Ia SN and showing the distinctive 
features of such a class of objects both in the visible and in the 
near-infrared.
Spectroscopically, SN~2000E appears as a ``normal'' Type Ia supernova, like
SN~1990N. 
We could constrain reddening [$E(B-V) \sim 0.5$ mag] and distance 
($\mu_{0} \sim 32.14$ mag) using a number of different methods. 
The bolometric luminosity curve of SN~2000E, which displays a bump 
at the epoch of the secondary near-infrared peak,
allows a determination of the $^{56}$Ni mass, amounting
to 0.9 M$_{\sun}$.
\end{abstract}


\section{Introduction}

Type Ia Supernovae (SNe Ia) are of great interest as
one of the most powerful tools for measuring the
Hubble constant and investigating the rate of expansion of the universe.
This is not only because of their large intrinsic luminosity,  
but is also due to the small dispersion in absolute 
$B$ magnitude at maximum light, which is less than 0.4 mag (Hamuy et al.\ 1996a).
The hope of further reducing this scatter has long driven
the search for empirical relations that could correlate 
the luminosity at maximum to morphological parameters of the light curves. 
The Phillips's relation and its subsequent modifications
(Phillips 1993, Hamuy et al.\ 1996a, Phillips et al.\ 1999),
use the decay in the $B$-band light curve from peak to 15 days
after peak, $\Delta m_{15} (B)$, and, although quite simple,
have proved effective in reducing the scatter around the Hubble flow to less 
than 0.14 mag, after reddening corrections. 
A more complex method based on light curve shapes in multiple optical 
passbands (i.e., multicolor light-curve shapes, or MLCS) has 
been developed by Riess, Press \& Kirshner (1996) and allows one   
to simultaneously estimate the SNe Ia luminosity and the amount of 
extinction, further reducing the scatter to only 0.12 mag.  
While probably more work is still needed, these studies clearly indicate
that light curve parametrization can be used to correct for the intrinsic
range in the absolute brightness of SNe Ia. 

The reliable use of SNe Ia for cosmological purposes requires an extremely
good calibration of the empirical relations. In this respect, the main problems
are the accuracy of independent distance determinations of nearby events
and the effects of reddening. In particular, the latter has 
highlighted the need for infrared observations, that
are less affected by interstellar dust. It has been already noted also that in the
near-infrared (NIR) SNe Ia appear generally quite homogeneous.
However, the coverage of NIR light 
curves is still much inferior to that achieved in the optical region. 
So far, only for a handful of events well sampled NIR light curves around peak
are available
(see, e.\ g., Meikle 2000, Krisciunas et al.\ 2000, and references therein;  
Krisciunas et al.\ 2001, Strolger et al.\ 2002,  Candia et al.\ 2003, Krisciunas et 
al.\ 2003). Nevertheless,
the absolute NIR magnitudes at 
 a given epoch past
maximum light show a dispersion of about $0.15$
mag for SNe with $0.87 < \Delta m_{15} (B) < 1.31$ (Meikle 2000).  
Therefore, NIR light curves could be used in principle to determine 
cosmological distances with an equal or probably greater accuracy
than the optical ones. 
In addition, combining NIR and optical photometry, it is possible to estimate 
the extinction due to the host galaxy (Krisciunas et al.\ 2000). 
At NIR wavelengths, the light curves of SNe Ia display a secondary peak after 
the optical maximum, whose shape and time of occurrence appear loosely related 
with the decline rate (Phillips et al.\ 2002, and references therein).
This suggests that other parameters characterizing the brightness evolution of
SNe Ia in a wide range of wavelengths may exist which still need an in-depth
exploration. A detailed knowledge of those parameters could not only allow one 
to find more reliable indicators of the absolute magnitudes at peak, 
but might also provide stronger constraints on the theoretical models
of SNe Ia progenitors and explosions. 

The use of empirical relations to minimize the uncertainty in the intrinsic 
luminosity of SNe Ia would greatly benefit by their fitting a well understood 
theoretical scenario. 
Several physical parameters [progenitor mass (if rotation is accounted for, 
see Piersanti et al. 2003) or thermal content, energy delivered during the 
explosion, mass of produced $^{56}$Ni, opacity of the photospheric layers] 
probably combine to shape the display of a SN Ia. 
However, the role of radiation transfer within the expanding structure
in correlating some of the observed features has been suggested as being 
crucial (Arnett 1982, Pinto \& Eastman 2000a, Pinto \& Eastman 2000b). This
is not unrelated to an in-depth understanding of the explosion mechanisms
and their effects on the final chemical composition and the physical conditions
within the SNe.

In the framework of the SWIRT (Supernova Watch-dogging InfraRed Telescope) 
project, a collaboration between the Observatories
of Collurania-Teramo, Rome and Pulkovo, 
we have been gathering accurate NIR data
for close-by SNe of all types since the late '90s. These can be usually complemented
by optical photometry from a few facilities operated by Italian Institutions. 
For SNe Ia, we aim to obtain a large 
data set of infrared light curves in order to fully characterize and parameterize
their behavior. Herein, we present the results of our follow-up of SN~2000E. The
paper is laid out as follows: in Sect.~\ref{odr}, observations and data reduction
procedures are summarized, in Sect.~\ref{res_an} our main results are described,
Sect.~\ref{disc} is devoted to discuss these results and in Sect.~\ref{concl}
we list our main conclusions.

\section{Observations and data reduction}
\label{odr}

%
%
\begin{figure}  
\epsscale{0.5}
\plotone{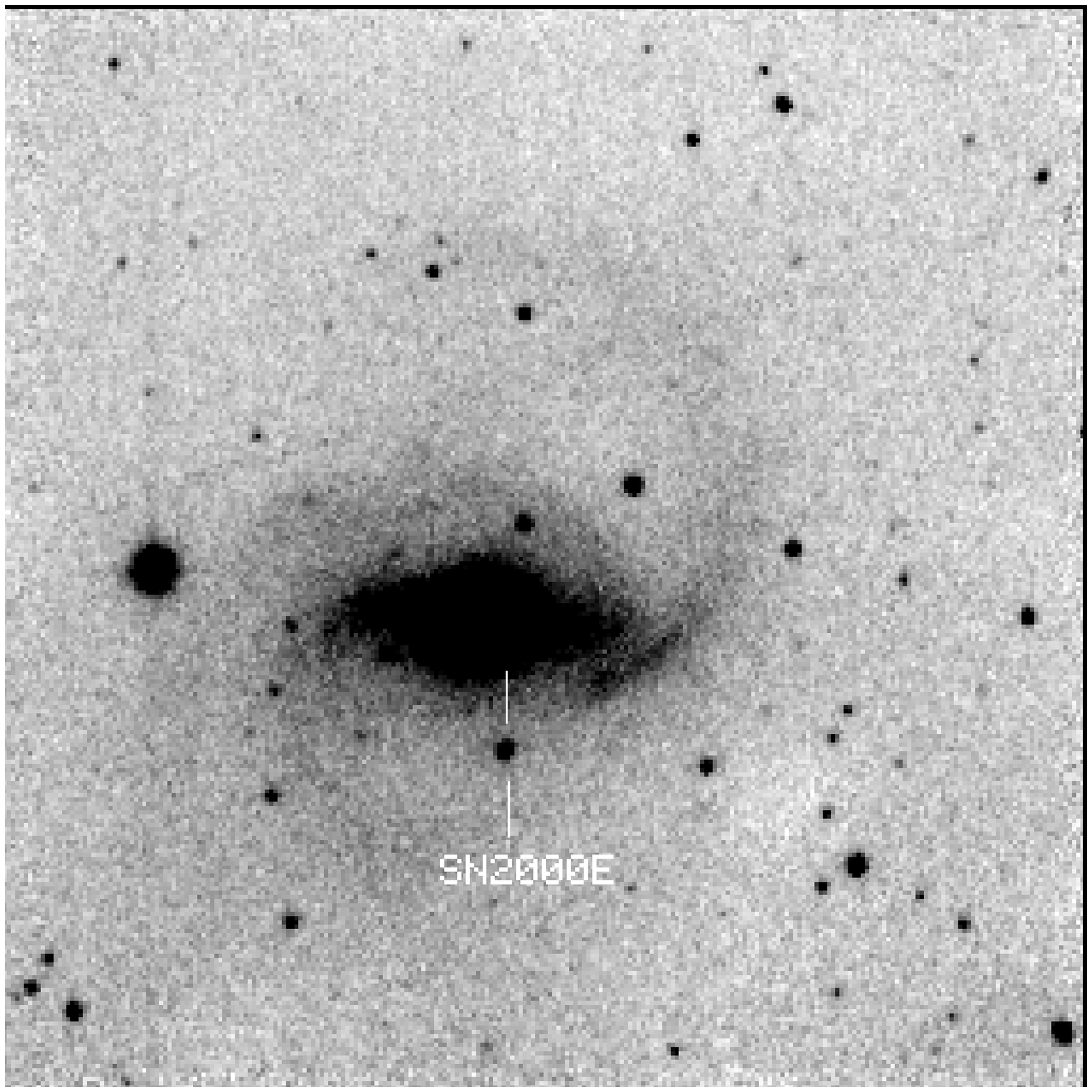}  
\caption{$K$-band image of SN~2000E in the galaxy NGC~6951 obtained at
 the AZT-24 on JD +576 (ph -1). North is up and East is to the left. 
 See Di Carlo et al.\ (2002) for the sequence of local photometric standards.  
\label{f:gal}}  
\end{figure}

SN~2000E was discovered on January 26.73 UT 
(JD 2451561; henceforth, JD+561) on CCD images obtained with the TNT telescope, located 
at the Teramo observatory (Italy), during routine observations of SN~1999el (Valentini 
et al.\ 2000). It was identified as a Type Ia by 
Turatto, Galletta \& Cappellaro  (2000) through spectral 
observations obtained with the 1.82-m telescope at Cima Ekar (Asiago, Italy) on January 
27.83 UT. It was located $6.3 \arcsec$ west and $26.7 \arcsec$ south of the nucleus of 
NGC 6951 (see Fig.~\ref{f:gal}),
an inclined ($i=42 \degr$) barred spiral galaxy hosting a type 2 Seyfert AGN
whose coordinates are 
$\alpha(2000.0)=20:37:13.97$ and $\delta(2000.0)=+66:06:19.87$. 
For a list of galaxy properties, see 
Kohno, Kawabe \& Vila-Vilar\'o (1999) and references therein.
The quoted distance is $24.1$ Mpc.  Although Valentini et al.\ (2000) report 
having obtained no 
images from December 18.78 UT (when the object does not 
appear on the frames) to the discovery day, 
SN~2000E was clearly detected at an epoch preceding the time of maximum.
The earliest record is claimed by
Evans \& Corso (2000), who found the object on $R$ and $I$ images taken
on January 24.03 UT.

The NIR observations were carried out using the AZT-24 telescope at
Campo Imperatore (Italy). 
On JD+691.62, we also obtained $JHK_{\rm s}$ images 
with ARNICA (Lisi et al.\ 1996) at the 3.58 m Telescopio Nazionale Galileo 
(TNG; La Palma, Canary Islands, Spain). \\
The AZT-24 is a 1.1 m Ritchey-Chr\'etien system 
equipped with the near-infrared camera SWIRCAM (D'Alessio et al.\ 2000), 
which is based on a 256$\times$256 HgCdTe PICNIC array.
The detector is sensitive to radiation in the spectral
range from 0.9 to 2.5 micron and allows a resolution of $1.04$ 
arcsec/pixel, for a total field-of-view of 4.4$\times$4.4 arcmin$^2$.
The available JHK filters are in the standard of the Johnson photometric system (1965). 
The scientific data were acquired collecting 5 dithered images of the
target and of a nearby sky location in each 
of the $J$, $H$, $K$ bands. 
The final scientific images were then obtained subtracting the sky 
background, flat-fielding, removing the bad pixels and
finally registering and combining the dithered exposures. 
The instrumental magnitudes were obtained via PSF (point-spread function) 
fitting photometry and the 
calibration to the standard system was performed through comparison with
a local sequence of field stars. Two of these stars were calibrated   
on $JHK_{s}$ images taken with the ARNICA 
camera at the TNG, 
using the standard star AS 40-5 (Hunt et al. 1998), as described 
in Di Carlo et al.\ (2002; see also their Fig. 1 for the local
sequence and the text for further details). 
The $JHK$ photometry of SN~2000E is listed in Table \ref{t:tab1}. The
uncertainties due to photon statistics and fitting are also indicated.

\begin{deluxetable}{ccccccccc}  
\tablecaption{NIR photometry of \protect\e\ obtained with SWIRCAM.
                                                                  \label{t:tab1}} 
\tablewidth{0pt}
\tablehead{
\colhead{Julian Day} & \colhead{Epoch} & \colhead{$J$} & \colhead{$\Delta J$} & \colhead{$H$} 
& \colhead{$\Delta H$} & \colhead{$K$} & \colhead{$\Delta K$}& \colhead{Telescope}  \\  
 \colhead{(2451000+)}& \colhead{(days)} & \colhead{(mag)} & \colhead{(mag)} & 
\colhead{(mag)} & \colhead{(mag)} & \colhead{(mag)} & \colhead{(mag)} & \colhead{} 
} 
\startdata
$  569.26 $ & $ -7.74  $ &  $  13.63 $ &  $  0.03 $ &  $  13.93 $ &  $  0.07 $ & $    -    $  &  $    -  $  &   $ AZT-24   $  \\  
$  570.28 $ & $ -6.72  $ &  $  13.60 $ &  $  0.01 $ &  $  13.74 $ &  $  0.06 $ & $  13.59  $  &  $  0.08 $  &   $ AZT-24   $  \\  
$  571.23 $ & $ -5.77  $ &  $  13.54 $ &  $  0.02 $ &  $  13.77 $ &  $  0.05 $ & $  13.52  $  &  $  0.06 $  &   $ AZT-24   $  \\  
$  574.27 $ & $ -2.73  $ &  $  13.47 $ &  $  0.02 $ &  $  13.88 $ &  $  0.07 $ & $  13.42  $  &  $  0.06 $  &   $ AZT-24   $  \\  
$  575.27 $ & $ -1.73  $ &  $  13.57 $ &  $  0.04 $ &  $  13.93 $ &  $  0.06 $ & $    -    $  &  $    -  $  &   $ AZT-24   $  \\  
$  576.22 $ & $ -0.78  $ &  $  13.58 $ &  $  0.03 $ &  $  13.86 $ &  $  0.07 $ & $  13.45  $  &  $  0.05 $  &   $ AZT-24   $  \\  
$  584.71 $ & $  7.71  $ &  $  14.51 $ &  $  0.02 $ &  $  14.05 $ &  $  0.07 $ & $    -    $  &  $    -  $  &   $ AZT-24   $  \\  
$  586.31 $ & $  9.31  $ &  $  14.64 $ &  $  0.06 $ &  $  14.13 $ &  $  0.08 $ & $  14.06  $  &  $  0.08 $  &   $ AZT-24   $  \\  
$  588.25 $ & $ 11.25  $ &  $  14.94 $ &  $  0.03 $ &  $  14.02 $ &  $  0.08 $ & $    -    $  &  $    -  $  &   $ AZT-24   $  \\  
$  599.69 $ & $ 22.69  $ &  $  14.79 $ &  $  0.06 $ &  $    -   $ &  $    -  $ & $  13.69  $  &  $  0.04 $  &   $ AZT-24   $  \\  
$  601.64 $ & $ 24.64  $ &  $  14.68 $ &  $  0.05 $ &  $  13.86 $ &  $  0.02 $ & $  13.53  $  &  $  0.04 $  &   $ AZT-24   $  \\  
$  623.57 $ & $ 46.57  $ &  $  15.46 $ &  $  0.06 $ &  $  14.50 $ &  $  0.06 $ & $  14.26  $  &  $  0.10 $  &   $ AZT-24   $  \\  
$  624.52 $ & $ 47.52  $ &  $  15.51 $ &  $  0.05 $ &  $     -  $ &  $    -  $ & $    -    $  &  $    -  $  &   $ AZT-24   $  \\  
$  636.66 $ & $ 59.66  $ &  $  16.17 $ &  $  0.06 $ &  $  15.13 $ &  $  0.07 $ & $  14.90  $  &  $  0.04 $  &   $ AZT-24   $  \\  
$  667.61 $ & $ 90.61  $ &  $  17.80 $ &  $  0.15 $ &  $  16.45 $ &  $  0.15 $ & $    -    $  &  $    -  $  &   $ AZT-24   $  \\  
$  689.62 $ & $ 112.62 $ &  $  18.94 $ &  $  0.15 $ &  $     -  $ &  $    -  $ & $    -    $  &  $    -  $  &   $ AZT-24   $  \\  
$  691.62 $ & $ 114.62 $ &  $ > 19.36 $ &  $    -  $ &  $  17.16 $ &  $  0.10 $ & $  16.74  $  &  $  0.09 $  &   $TNG/ARNICA$  \\  
$  703.58 $ & $ 126.58 $ &  $    -   $ &  $    -  $ &  $  17.58 $ &  $  0.20 $ & $    -    $  &  $    -  $  &   $ AZT-24   $  \\  
\enddata
\end{deluxetable}  

Most of the optical images were collected with a 512x512 Tektronics optical
camera
mounted at the focal plane of the 0.72 m TNT telescope at the Teramo
Observatory.
During several nights spanning January to March 2000,  
we obtained different sets of images of the supernova field
in the $BVRI$ bands. These images were processed using
standard bias-subtraction and flat-field normalization techniques and the
instrumental
magnitudes were obtained via PSF photometry on each image after
averaging four or more measurements available per night. 

Optical images at $UBVRI$ were also taken with 
the 1.82-m Copernicus telescope (henceforth CT) operated by the Osservatorio Astronomico di
Padova and sited in Asiago, Cima Ekar (Italy), equipped with AFOSC. The
instrument is a focal reducer type spectrograph/camera with a 1K
$\times$ 1K Site Thinned CCD ($24\mu$m). The scale of 0.473 arcsec/pixel
gives a field of view of $8 \times 8$ arcmin$^{2}$.

Some $UBVRI$ photometry was obtained with the Optical Imager Galileo
(OIG) mounted on TNG and equipped with a mosaic of two thinned, back-illuminated 
EEV42-80 CCDs with $2048\times 4096$ pixels each (pixel
size $13.5$ $\mu$m; pixel scale in 2x2 binned mode $0.144$ arcsec/pixel). Some 
of these nights were photometric and were used to calibrate the
local sequence around the SN 
which in turn was used for calibrating non-photometric nights
at all of the telescopes.
The instrumental color equations were obtained through observations of
stars in the standard fields of Landolt (1992).

Table \ref{t:tab2} lists the results of optical photometry for \protect\e,
also shown in Fig.~\ref{f:LC}.  All UBVRI magnitudes have been standardized to the Johnson-Cousins filter system (Bessel 1979). 
Quoted errors are the uncertainties arising from the photon statistics and 
fitting.  
As can be seen in Fig.~\ref{f:LC}, there are small discrepancies at $B$-band
between data points from the TNT and the CT.
The effect can be described to first order as an offset between the
two sets of measurements of $\sim 0.1$ mag, at least during the period from JD+570
to JD+590.
The small systematic differences are not unexpected and probably
arise owing to the non-stellar nature of SN spectra which enhances filter and detector 
differences between instruments (see, e.\ g., Stritzinger et al.\ 2002). Although
spectro-photometric methods to correct for these effects are currently being attempted
(see, e.\ g., Stritzinger et al.\ 2002, Candia et al.\ 2003, Krisciunas et al.\ 2003),
it is not one of our aims to perform such corrections. 
Anyway, these discrepancies
between measurements are taken into account in the following. 

As reported by Di Carlo et al.\ (2002), the local sequence used here for differential
photometry includes 3 stars among those selected by Vink\'o et al. (2001) to
perform their own differential photometry of SN~2000E. The magnitudes of the 3 objects
agree within $0.05$ with 2 exceptions in the $R$ and $I$ band (where, however, the
difference is $\sim 0.1$ mag), confirming the validity of
both our and their calibrations. 

We compared the $BVRI$ magnitudes given by  Vink\'o et al.\ (2001) in their Tab.~3 
with the values obtained by spline fits to our data (see
Sect.~\ref{bolo:sec}). This allows one to circumvent the shortage of 
overlapping observations in their and our datasets.
The differences (in the sense their values minus ours) are summarized
in Table~\ref{vin:noi}.
Between JD+572 and JD+589, when we have  a good time
coverage, $V$ and $R$ coincide within $\sim 0.1$ mag (often within $\sim 0.05$ mag).
The same is true for $I$, although we lack points between JD+575 and JD+586 
(just the spline curve); at JD+576, however, $I$ differs by $\sim 0.2$ mag, 
but the quality of their $I$-band datum seems lower (see their Fig.~4).
Between JD+656 and JD+667, $BVR$ differ by $\sim 0.1-0.3$ mag, yet we lack any
measurements between JD+612 and JD+690  and the quality of spline interpolation over such a large interval is then poor.  After JD+690, $V$ again coincides 
within $\sim 0.05 - 0.1$ mag and the only comparable point at $R$ (JD+696) is 
somewhat discordant ($\sim 0.4$ mag). Also in this case, inspection of their 
Fig.~3 suggests that 
the value given by Vink\'o et al.\ (2001) may be somewhat less accurate. 
Part of the differences may be due to the different techniques employed: 
whereas they perform aperture photometry, we adopted a PSF-fitting 
photometry.

\begin{deluxetable}{ccccccccccccc}
\tabletypesize{\scriptsize}
\rotate
\tablecaption{$UBVRI$ (Johnson-Cousins) photometry for 
SN~2000E obtained with different instruments.
                                                                                  \label{t:tab2}}    
\tablewidth{0pt}
\tablehead{
\colhead{Julian Day} & \colhead{Epoch} &\colhead{$U$} & \colhead{$\Delta U$} & \colhead{$B$} & 
\colhead{$\Delta B$} & \colhead{$V$} & \colhead{$\Delta V$} & \colhead{$R$} & 
\colhead{$\Delta R$} & \colhead{$I$} & \colhead{$\Delta I$} & 
\colhead{Telescope} \\ 
\colhead{(2451000+)}& \colhead{(days)} & \colhead{(mag)} & \colhead{(mag)} & \colhead{(mag)} & 
\colhead{(mag)} & \colhead{(mag)} & \colhead{(mag)} & \colhead{(mag)} & 
\colhead{(mag)} & \colhead{(mag)} & \colhead{(mag)}&  \colhead{}
}
\startdata                                
$ 561.25 $& $ -15.75 $ &$    -  $ &$  -   $ &$ 17.84 $& $ 0.05 $&$    -   $ & $    -   $  & $ 16.73  $  & $ 0.13  $  & $ 16.60  $   & $ 0.20  $  & $ TNT  $ \\ 
$ 563.30 $& $ -13.70 $ &$    -  $ &$  -   $ &$ 16.77 $& $ 0.02 $&$    -   $ & $    -   $  & $ 15.63  $  & $ 0.02  $  & $ 15.42  $   & $ 0.03  $  & $ TNT  $ \\ 
$ 570.27 $& $ -6.73  $ &$    -  $ &$  -   $ &$ 14.69 $& $ 0.01 $&$ 14.170 $ & $  0.009 $  & $ 13.910 $  & $ 0.004 $  & $   -    $   & $  -    $  & $ TNT  $ \\ 
$ 571.25 $& $ -5.75  $ &$    -  $ &$  -   $ &$ 14.47 $& $ 0.02 $&$ 14.04  $ & $  0.03  $  & $ 13.768 $  & $ 0.004 $  & $   -    $   & $  -    $  & $CT$ \\ 
$ 572.24 $& $ -4.76  $ &$    -  $ &$  -   $ &$ 14.520$& $ 0.004$&$ 14.02  $ & $  0.02  $  & $ 13.760 $  & $ 0.006 $  & $   -    $   & $  -    $  & $ TNT  $ \\ 
$ 574.27 $& $ -2.73  $ &$    -  $ &$  -   $ &$ 14.430$& $ 0.004$&$ 13.900 $ & $  0.008 $  & $ 13.660 $  & $ 0.008 $  & $   -    $   & $  -    $  & $ TNT  $ \\ 
$ 574.32 $& $ -2.68  $ &$    -  $ &$  -   $ &$ 14.35 $& $ 0.03 $&$ 13.88  $ & $  0.02  $  & $ 13.72  $  & $ 0.05  $  & $ 13.56  $   & $ 0.05  $  & $CT$ \\ 
$ 575.29 $& $ -1.71  $ &$14.495 $ &$0.015 $ &$ 14.319$& $ 0.003$&$ 13.983 $ & $  0.002 $  & $ 13.602 $  & $ 0.007 $  & $ 13.510 $   & $ 0.005 $  & $CT$ \\ 
$ 579.25 $& $  2.25  $ &$    -  $ &$  -   $ &$ 14.42 $& $ 0.03 $&$ 13.830 $ & $  0.006 $  & $ 13.580 $  & $ 0.008 $  & $   -    $   & $  -    $  & $ TNT  $ \\ 
$ 580.25 $& $  3.25  $ &$    -  $ &$  -   $ &$ 14.44 $& $ 0.01 $&$ 13.840 $ & $  0.007 $  & $ 13.600 $  & $ 0.006 $  & $   -    $   & $  -    $  & $ TNT  $ \\ 
$ 581.30 $& $  4.30  $ &$    -  $ &$  -   $ &$ 14.47 $& $ 0.01 $&$ 13.860 $ & $  0.007 $  & $ 13.62  $  & $ 0.01  $  & $   -    $   & $  -    $  & $ TNT  $ \\ 
$ 583.30 $& $  6.30  $ &$    -  $ &$  -   $ &$ 14.58 $& $ 0.05 $&$ 13.920 $ & $  0.008 $  & $ 13.700 $  & $ 0.006 $  & $   -    $   & $  -    $  & $ TNT  $ \\ 
$ 584.70 $& $  7.70  $ &$14.957 $ &$0.032 $ &$ 14.569$& $ 0.007$&$ 13.95  $ & $  0.06  $  & $ 13.81  $  & $ 0.04  $  & $ 13.83  $   & $ 0.08  $  & $CT$ \\ 
$ 585.30 $& $  8.30  $ &$    -  $ &$  -   $ &$ 14.68 $& $ 0.08 $&$ 14.000 $ & $  0.005 $  & $ 13.810 $  & $ 0.005 $  & $   -    $   & $  -    $  & $ TNT  $ \\ 
$ 585.30 $& $  8.30  $ &$14.946 $ &$0.006 $ &$ 14.541$& $ 0.004$&$ 13.960 $ & $  0.007 $  & $ 13.81  $  & $ 0.04  $  & $ 13.84  $   & $ 0.03  $  & $CT$ \\ 
$ 589.30 $& $  12.30 $ &$     - $ &$  -   $ &$ 14.960$& $ 0.007$&$ 14.240 $ & $  0.006 $  & $ 14.120 $  & $ 0.008 $  & $   -    $   & $  -    $  & $ TNT  $ \\ 
$ 593.50 $& $  16.50 $ &$     - $ &$  -   $ &$ 15.51 $& $ 0.05 $&$ 14.480 $ & $  0.008 $  & $ 14.25  $  & $ 0.02  $  & $   -    $   & $  -    $  & $ TNT  $ \\ 
$ 597.30 $& $  20.30 $ &$     - $ &$  -   $ &$ 15.850$& $ 0.008$&$ 14.640 $ & $  0.009 $  & $ 14.260 $  & $ 0.008 $  & $   -    $   & $  -    $  & $ TNT  $ \\ 
$ 599.30 $& $  22.20 $ &$     - $ &$  -   $ &$ 16.040$& $ 0.007$&$ 14.710 $ & $  0.003 $  & $ 14.300 $  & $ 0.003 $  & $   -    $   & $  -    $  & $ TNT  $ \\ 
$ 604.50 $& $  27.50 $ &$     - $ &$  -   $ &$ 16.52 $& $ 0.01 $&$ 14.930 $ & $  0.004 $  & $ 14.340 $  & $ 0.003 $  & $   -    $   & $  -    $  & $ TNT  $ \\ 
$ 604.50 $& $  27.50 $ &$ 17.40 $ &$ 0.03 $ &$ 16.668$& $ 0.008$&$ 14.999 $ & $  0.006 $  & $ 14.375 $  & $ 0.003 $  & $ 13.812 $   & $ 0.004 $  & $ TNG  $ \\ 
$ 605.50 $& $  28.50 $ &$     - $ &$  -   $ &$   -   $& $   -  $&$ 15.11  $ & $  0.01  $  & $ 14.46  $  & $ 0.02  $  & $   -    $   & $  -    $  & $ TNG  $ \\ 
$ 612.30 $& $  35.30 $ &$     - $ &$  -   $ &$ 17.10 $& $ 0.02 $&$ 15.440 $ & $  0.009 $  & $ 14.76  $  & $ 0.02  $  & $   -    $   & $  -    $  & $ TNT  $ \\ 
$ 690.50 $& $ 113.50 $ &$ 19.26 $ &$ 0.09 $ &$ 18.34 $& $ 0.03 $&$ 17.57  $ & $  0.03  $  & $ 17.49  $  & $ 0.04  $  & $ 17.42  $   & $ 0.02  $  & $ TNG  $ \\ 
$ 698.54 $& $ 121.54 $ &$     - $ &$  -   $ &$ 18.44 $& $ 0.09 $&$ 17.75  $ & $  0.14  $  & $ 17.70  $  & $ 0.09  $  & $ 17.53  $   & $ 0.13  $  & $CT$ \\ 
$ 699.55 $& $ 122.55 $ &$     - $ &$  -   $ &$ 18.42 $& $ 0.05 $&$ 17.76  $ & $  0.09  $  & $ 17.65  $  & $ 0.13  $  & $   -    $   & $  -    $  & $CT$ \\ 
$ 781.58 $& $ 204.58 $ &$     - $ &$  -   $ &$ 19.668$& $ 0.017$&$ 19.194 $ & $  0.019 $  & $ 19.677 $  & $ 0.017 $  & $   -    $   & $  -    $  & $TNG$ \\ 
$ 795.39 $& $ 218.39 $ &$     - $ &$  -   $ &$ 20.00 $& $ 0.22 $&$ 19.71  $ & $  0.61  $  & $   -    $  & $   -   $  & $   -    $   & $  -    $  & $CT$ \\ 
\enddata
\end{deluxetable} 
%
%
\begin{deluxetable}{ccccc}
\tablecaption{Comparison between the optical photometry of SN~2000E from
	 Vink\'o et al.\ (2001) and that from this work. The quoted
	 values are the average of their $BVRI$ mag minus ours in the
	 indicated time span. 
	 \label{vin:noi}}
	 \tablewidth{0pt}
	 \tablehead{
	 \colhead{Epoch} & \colhead{$<\Delta B>$} & \colhead{$<\Delta V>$} &
	 \colhead{$<\Delta R>$} & \colhead{$<\Delta I>$} \\
	 \colhead{} & \colhead{(mag)} & \colhead{(mag)} & \colhead{(mag)} &
	  \colhead{(mag)} \\
	 }
	 \startdata
From JD+572 to JD+589 & - & $-0.07 \pm 0.05$ & $-0.05 \pm 0.05$ & $-0.02 \pm 0.10$ \\
From JD+656 to JD+667 & $-0.12 \pm 0.07$ & $-0.25 \pm 0.08$ & $-0.19 \pm 0.10$ & - \\
From JD+696 to JD+727 & - & $-0.04 \pm 0.04$ & $-0.38 \pm 0.01$ & - \\
\enddata
\end{deluxetable}

\section{Results and analysis}
\label{res_an}

\subsection{Light curves}
\label{LC:sec}

\begin{figure}
\includegraphics[height=18cm]{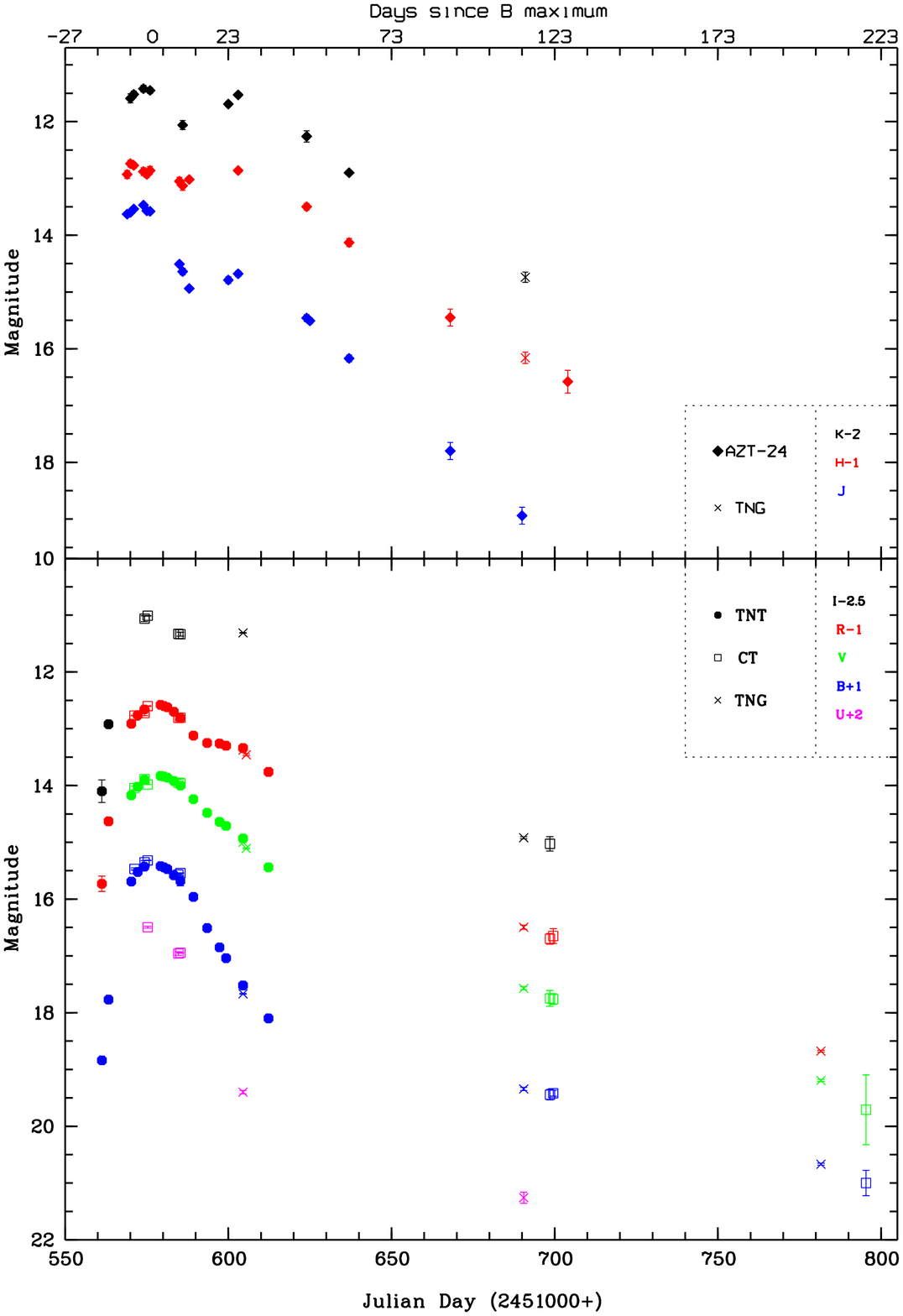}  
\caption{Optical and infrared light curves of SN~2000E. Errors marked
 by the bars are those indicated in Tab.~\ref{t:tab1} and \ref{t:tab2}.
\label{f:LC}}
\end{figure}

The light curves of \e\ are shown in Fig.~\ref{f:LC}. 
As can be seen there and in Tab.~\ref{t:tab2},
we began our observations through the 
$B$, $R$ and $I$ filters about 16 days before maximum $B$ brightness. 
However, the detection by Evans \& Corso (2000) sets the rise
time to $B_{\rm max}$, to at least 17--18 days.
The NIR observations began $\sim 7$ days before $B_{\rm max}$, 
allowing a good coverage around that epoch at these wavelengths, as well.
Currently only a few studies of SNe Ia, almost all quite recent,
cover the epoch of maximum NIR light (preceding $B_{\rm max}$ a few
days). This is instrumental in characterizing the LCs at all wavelengths,
even more so if we consider that, e.\ g., emission at $K$ suffers from
a reddening which is a factor $\sim 10$ less than at $V$, making the
NIR suitable for a more accurate determination of absolute brightness,
decreasing the uncertainties for the estimate of distances (Meikle 2000).

The time and magnitude at maximum light were estimated in each band
(excepted $U$) via 
third-order polynomial fits to the data around the peaks.
The results are listed in Tab.~\ref{peak_date}. 
Since the fits to the 
optical light curves may be sensitive to systematic differences
between photometry from different instruments already pointed out at $B$, 
we gave the same weight to all data assigning them a minimum error of 0.05 mag. 
For the $B$-band, we chose to
limit the data set to the measurements from the TNT only. However, the values
obtained shifting the latter ones by 0.1 mag to have them agree with
the photometric points from the CT are also indicated. 
For $I$, we give a conservative estimate of the error because of the
shortage of data points (implying a less reliability of the results). 
Finally we could not fit the main peak of $H$ with the polynomial curve, so 
we performed instead a spline fit (note that spline fits to $J$ and $K$ yield 
roughly the same values as the polynomial fits). 
From the estimated magnitude at $B$ maximum and a linear fit to the 
declining part of the curve (for
the TNT data set), we derive a decline rate at $B$, as defined by Phillips 
(1993), $\Delta m_{15}(B) = 0.94 \pm 0.05$.  
Since the discrepancies between the two sets of optical data (TNT and CT) can 
be described as a constant offset, using the one with the best time coverage 
should result in an accurate enough estimate of the decline rate regardless 
of the photometric differences.
According to Phillips et al.\ (1999), $\Delta m_{15}(B)$ is $\sim 1.1$ for a
typical SN Ia and it has been reported to range from $\sim 0.75$
(SN~1999aa; Krisciunas et al.\ 2000) to $\sim 1.94$ (SN~1999da; 
Krisciunas et al.\ 2001). Hence, SN~2000E may be considered as a
``slow decliner'' and 
 is therefore very likely overluminous, as predicted by 
the empirical correlation found between $\Delta m_{15}(B)$ and the
intrinsic magnitudes of SNe Ia (Phillips 1993; Hamuy et al.\ 1996a; Phillips 
et al.\ 1999). 

\begin{figure}
\includegraphics[height=12cm,angle=270]{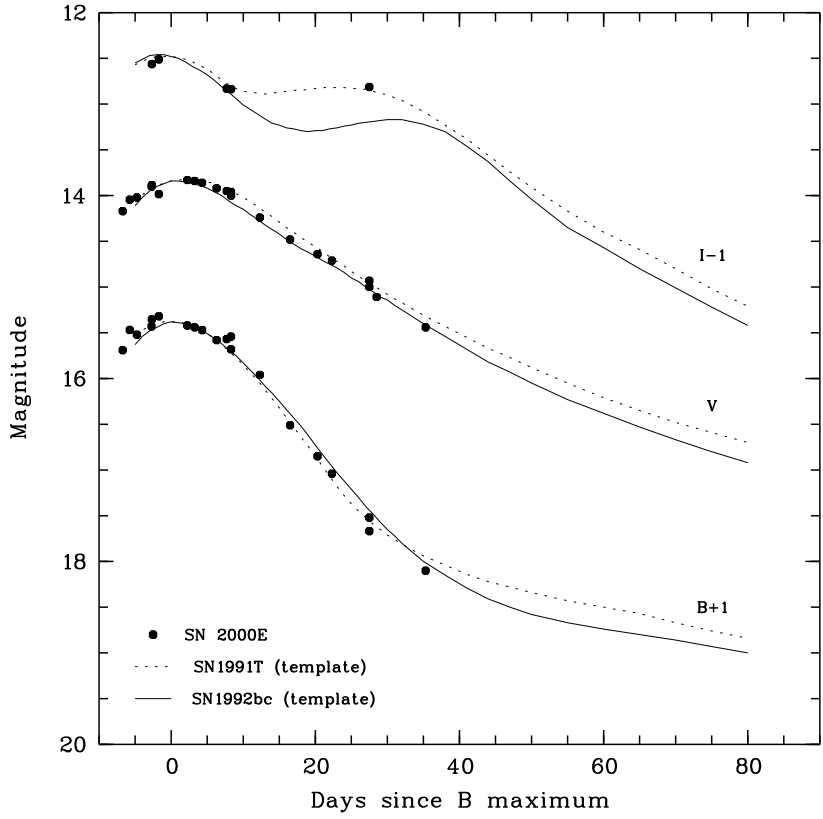}  
\caption{$BVI$ light curves of SN~2000E compared with
the template light curves of SN~1992bc ($\Delta m_{15} (B)=0.87$ mag) and SN~1991T
($\Delta m_{15} (B)=0.94$ mag), from Hamuy et al.\ (1996b).
\label{templ:f}}
\end{figure}

We also compare the optical light curves with the templates of Hamuy et al.\ 
(1996b). In Fig.~\ref{templ:f}  our LCs at $BVI$ are overlaid with those of 
the slow decliners SN~1992bc ($\Delta m_{15}(B) = 0.87$) and SN~1991T
($\Delta m_{15}(B) = 0.94$). We note that at $B$ and $V$ the LCs of SN~2000E 
are well approximated by the two templates, confirming our estimate of $\Delta 
m_{15}(B)$. The few data points at $I$ are better represented by the template 
of SN~1991T. The differences between LCs at $I$ may suggest that a single
parameter, e.\ g.\ $\Delta m_{15}(B)$, cannot actually describe the brightness 
evolution of SNe Ia at all wavelengths, a conclusion depending only on few 
points.

%
%
\begin{deluxetable}{ccccc}
\tablecaption{Optical and NIR magnitudes at peak along with the time 
	 of maximum light. \label{peak_date}}
	 \tablewidth{0pt}
	 \tablehead{
	 \colhead{Band} & \colhead{Julian Day} & \colhead{Magnitude}  \\
	 \colhead{ } & \colhead{at maximum} & \colhead{at maximum}\\
	 }
	 \startdata
$B$\tablenotemark{a} & $577.0 \pm 0.5$ & $14.39 \pm 0.03$ \\
$B$\tablenotemark{b} & $577.1 \pm 0.4$ & $14.30 \pm 0.02$ \\
$V$ & $579.1 \pm 0.7$ & $13.84 \pm 0.02$ \\
$R$ & $578.2 \pm 0.5$ & $13.59 \pm 0.02$ \\
$I$ & $577.4 \pm 1.0$ & $13.48 \pm 0.06$ \\
$J$ & $573.2 \pm 0.1$ & $13.48 \pm 0.01$ \\
$H$\tablenotemark{c} & $571.0 \pm 1$ & $13.76 \pm 0.05$ \\
$K$ & $574.4 \pm 1.5$ & $13.43 \pm 0.05$ \\
\enddata
\tablenotetext{a}{From TNT data only.}
\tablenotetext{b}{Shifting the TNT data onto the CT data.}
\tablenotetext{c}{From spline fits to the data.}
\end{deluxetable}

\begin{figure}
\includegraphics[height=12cm,angle=270]{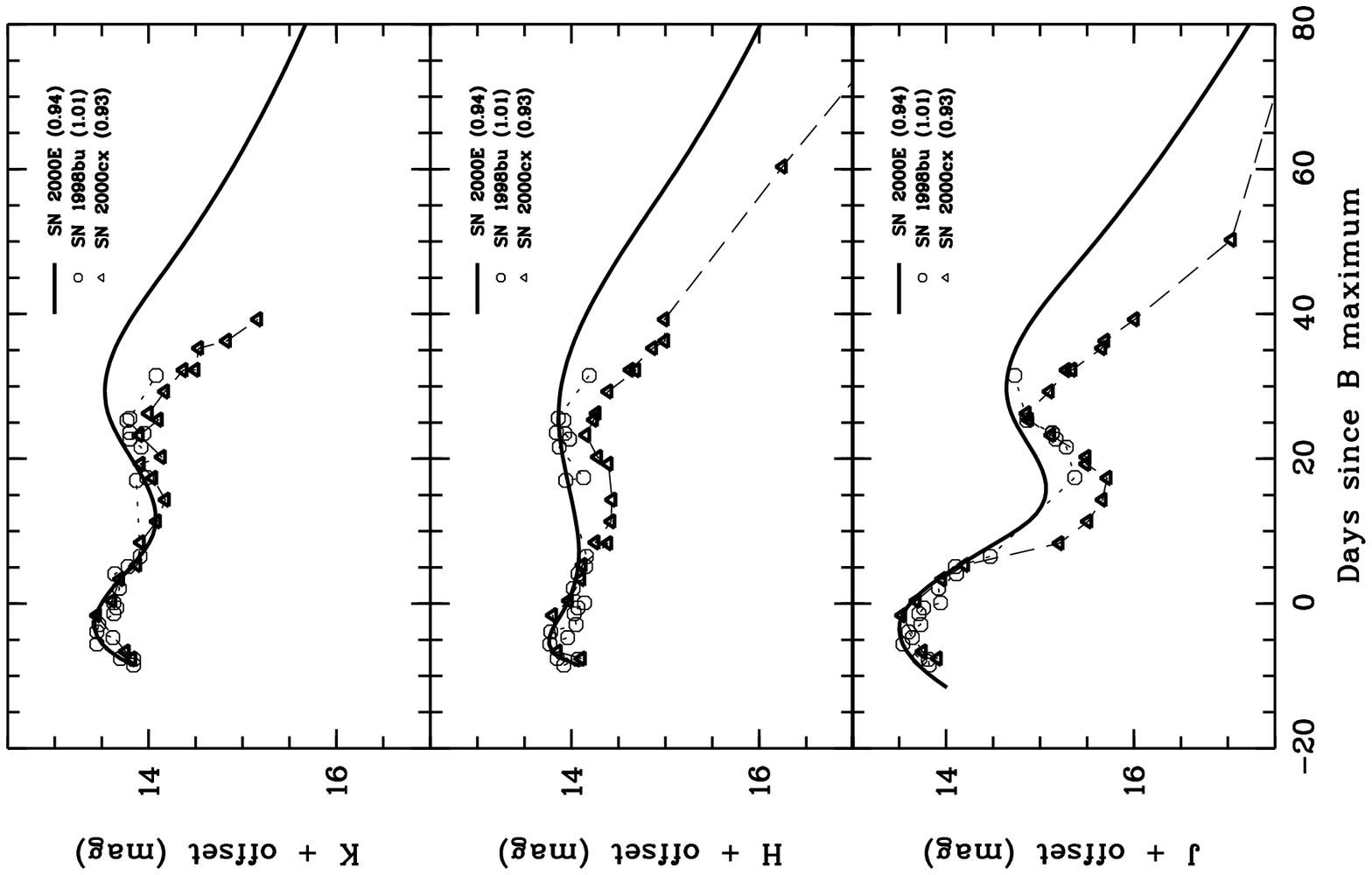} 
\caption{$JHK$ light curves (from spline fits to data) 
of SN~2000E (solid line) compared with
the light curves of SN~1998bu (Meikle 2000)
and SN~2000cx, a peculiar slow decliner (Candia et al.\ 2003). Data points are 
connected with dotted (SN 1998bu) and dashed (SN 2000cx) lines for the sake of 
clarity. 
\label{nirconf:f}}
\end{figure}

In the NIR, the LCs display the  characteristic double peak which is also
noticeable at $R$ and $I$ (see Fig.~\ref{f:LC}). The earlier (and brighter) peak 
occurs at around JD+571--574, 
3 to 6 days before $B_{\rm max}$, whereas the secondary one
occurs roughly on JD$+607$, i.\ e., 30 days past $B_{\rm max}$. 
A comparison between the NIR light curves of SN~2000E and those
of SN~1998bu (Meikle 2000) and 
SN~2000cx (Candia et al.\ 2003), with $\Delta m_{15} (B) = 1.01$ and 0.93,
respectively, is shown in Fig.~\ref{nirconf:f}, 
which further emphasizes the inadequacy of a uniparametric characterization 
of LCs. All share a well developed 
secondary peak, although it seems to occur later for SN~2000E at $K$ and 
probably also at $H$. The LC of SN~2000E 
is extremely flat between maxima at $H$ and the minima are the shallowest at 
both $H$ and $J$.  SN~2000E appears also
the one with the slowest decline in the NIR of the three objects 
(SN~2000cx is however a peculiar slow decliner). 

%
%

%
%

%
%

\subsection{Color curves}
\label{col:sec}

\begin{figure}
\epsscale{0.8}
\includegraphics[height=18cm,angle=270]{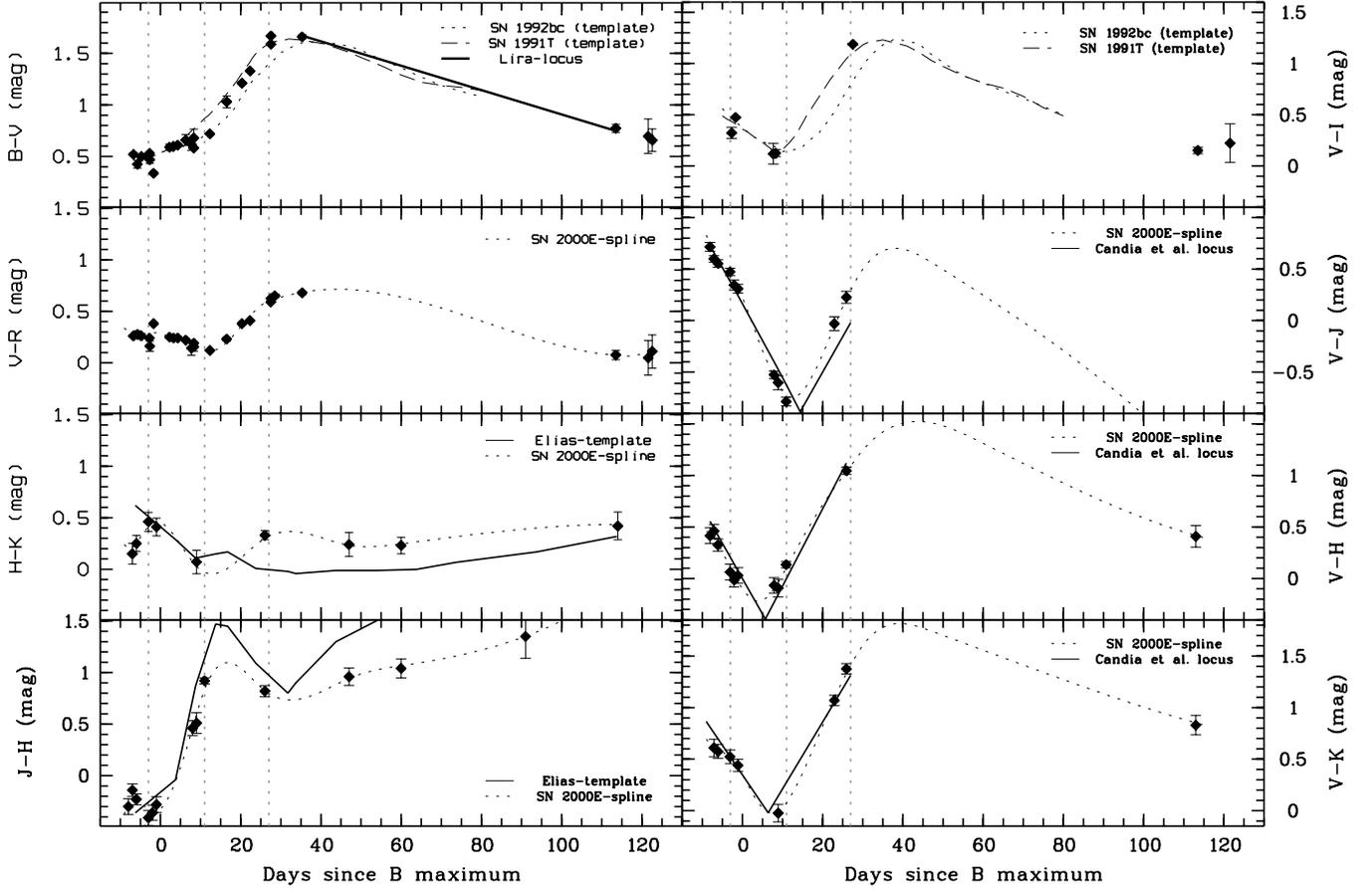}  
\caption{Optical and infrared colors of SN~2000E. 
 The locus of Lira (solid line) is shown in the upper left 
 box after shifting to account for reddening.
 For the sake of comparison, the $B-V$ and $V-I$ color templates
 of SN~1991T (dashed line) and SN~1992bc (dotted line)
 are drawn. The $H-K$ and $J-H$ templates from Elias et al.\ (1981) are also
 shown (solid line in the corresponding boxes).
 The $V$ minus NIR colors for slow decliners (solid line; see text)
 are also superimposed on the data points (dotted line: color curves
 from spline fits) after shifting to account for reddening. 
 Vertical dotted lines mark the time of the earlier NIR maximum,
 the minimum between NIR peaks and the secondary NIR maximum; these
 have been drawn to show the optical color behavior during such
 phases of the NIR light curves (see text).
\label{f:col}}
\end{figure}

Optical and NIR colors are plotted in Fig.~\ref{f:col}. 
In the upper left panel we compare
$B-V$ with template curves given by Hamuy et al. (1996b).
Again we find that the $B-V$ curve of SN~2000E has  a similar shape as 
those of SN~1992bc and SN~1991T. 
SNe Ia with $0.85 \la \Delta m_{15} (B) \la 1.90$ and little or no
reddening from their host galaxy, show a very uniform $B-V$ color evolution at 
late epochs 
past maximum light (Lira 1995, Riess et al.\ 1996). Such color uniformity
has proven useful as an indicator of host galaxy reddening, and has been 
important in revising the empirical relations characterizing LCs.
Lira (1995) found that the $B-V$ colors of Type Ia
SNe from 30 to 90 days past $V$ maximum evolve in the same manner and derived 
a relation to describe
the intrinsic $B-V$ colors (Phillips et al.\ 1999). During this time interval
(JD+609 to JD+669) we have only one data point, so we derived $B-V$ colors also
from template fits to the LCs at $B$ and $V$.
We find that the offset between the Lira relation and the $B-V$ 
colors of SN~2000E is consistent with the reddening (see Sect.~\ref{Red}).

Also in the $V$ minus NIR color evolution, spectroscopically normal SNe Ia
show an impressive uniformity in their intrinsic colors around $B$ maximum 
light. Krisciunas et al.\
(2000, 2001) found that for mid-range decliners these are uniform from 9 days
before maximum to 27 days after maximum. 
Candia et al.\ (2003) construct 
the same loci but for slow decliners. Using weighted fits to the LCs at
$VJHK$, after deriving the $V-J$, $V-H$ and $V-K$ colors from JD+569 to JD+604,
we found that the differences between them and the relations given by Candia et al.\
(2003; see also their Fig.~13) are fully consistent with reddening (see Sect.~\ref{Red}).
This result, and the agreement of $B-V$ colors with the Lira relation
after dereddening, both shown in Fig.~\ref{f:col}, not only allow
an estimate of extinction, but also confirm that SN~2000E is a slowly 
declining
Type Ia SN without peculiar behavior.

\begin{figure}
\epsscale{0.8}
\plotone{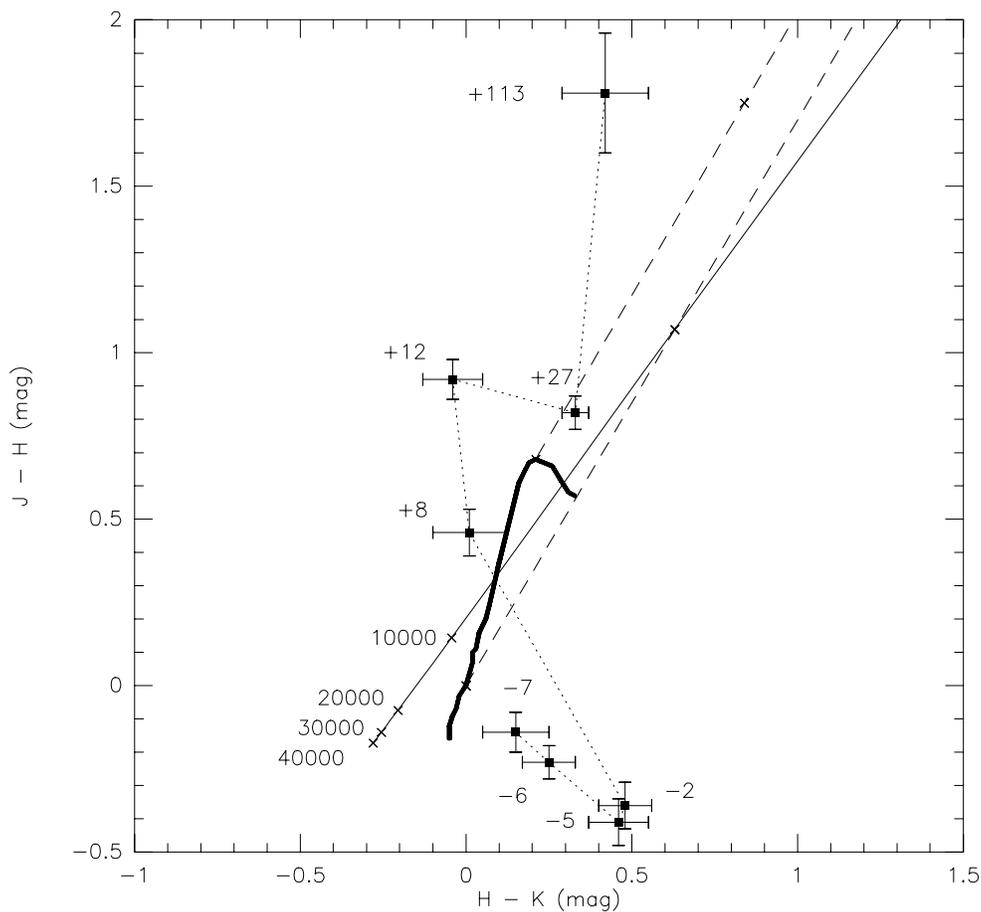}
\caption{Color-color diagram ($H-K$ vs.\ $J-H$) showing
 the time evolution of the NIR emission from SN~2000E.
 The filled squares (with error bars) mark the measured colors,
 which evolve along the dotted line from JD$+570$ 
 to JD$+690$ (the epoch is indicated near each data point
 in days since $B_{\rm max}$).
 The solid black line reproduces the main sequence locus
 from O6--8 to M8 stars (according to the colors given
 by Koornneef 1983), whereas the dashed lines correspond to 
 reddening according to the law given by Cardelli et al. (1989).
 The straight line indicates blackbody colors with
 crosses at T=40000, 30000, 20000 and 10000 K.
\label{f:nir-col-col}}
\end{figure}

\begin{figure}
\epsscale{1.0}
\plotone{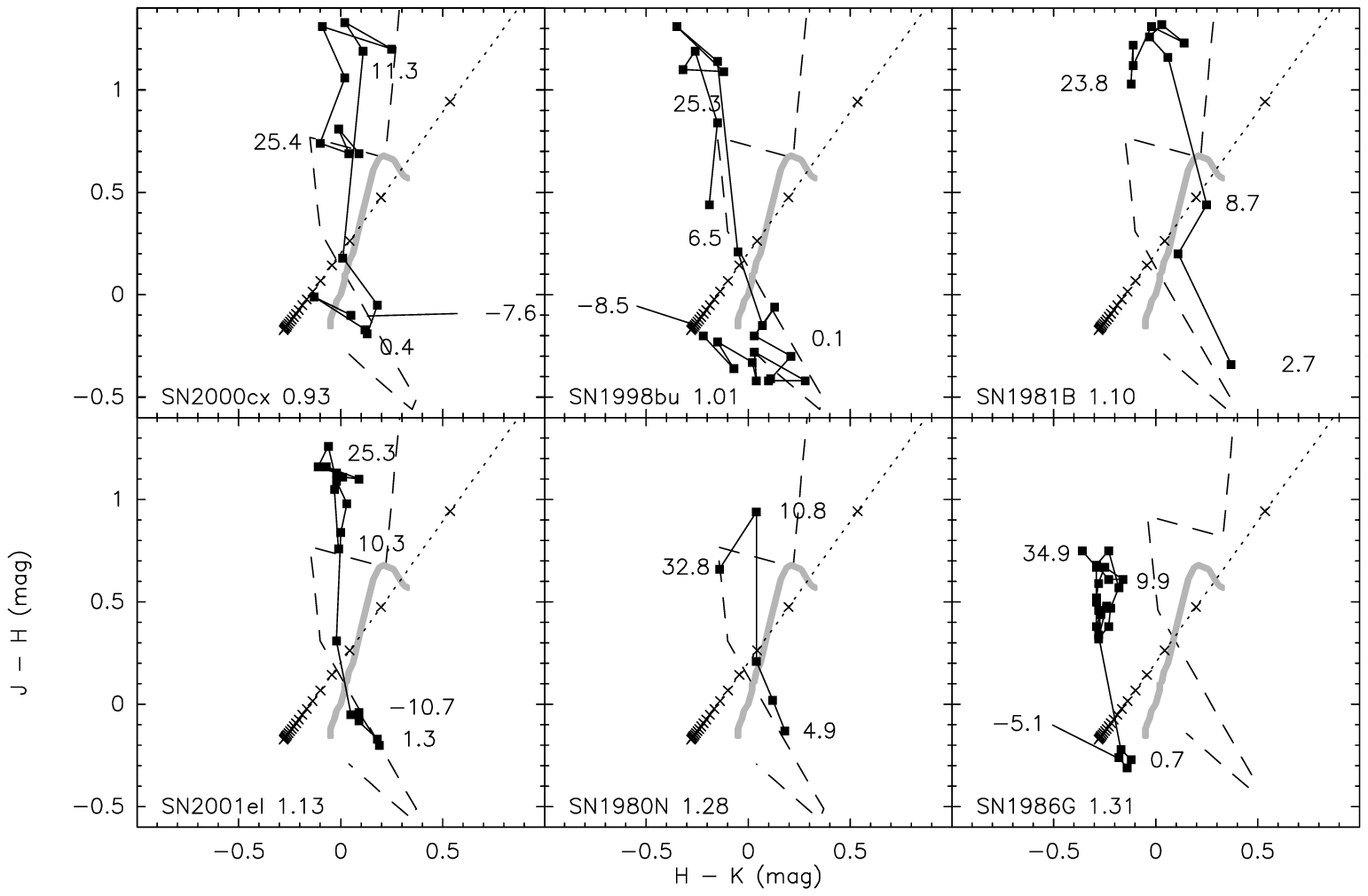}
\caption{Comparison of color-color diagrams ($H-K$ vs.\ $J-H$) 
 for 6 Type Ia SNe in a range of $\Delta m_{15} (B)$,
 overlaid with SN~2000E (dashed line).
 At the bottom of each box, SN and $\Delta m_{15} (B)$ 
 are indicated.
 The dotted line is the blackbody locus, the grey one is
 the main sequence locus. For each supernova,
 some data points are labeled with their phase (in
 days): first NIR
 observation, around $B_{\rm max}$ (phase $\sim 0$ d), around minimum
 between NIR peaks (phase $\sim 10 $d) and around secondary NIR peak
 (phase $\sim 25 $d). All colors have been dereddened.
\label{f:nir-col-comp}}
\end{figure}

It is interesting to follow the evolution of NIR
colors in between the two peaks. 
As can be checked in Fig.~\ref{f:col}, $J-H$ and $H-K$ display
a different behavior: $H-K$ appears to decrease from $\sim 0.5$ to
$\sim 0.1$ mag after the first peak, until the minimum between the two peaks
is reached. Then, it increases to $\sim 0.3$ mag 
at the epoch of the secondary peak. On the other hand, $J - H$ increases
from $\sim -0.4$ mag to $\sim 1$ mag after the earlier maximum, 
until the minimum between the two peaks is reached, then it slowly
fades to $\sim 0.8$ mag before the secondary maximum. Fig.~\ref{f:LC}
indicates that this essentially reflects the different main-peak-to-dip 
height, which is smaller in $H$ than in $J$ and $K$.
The time evolution of NIR colors is fully consistent with that envisaged 
by Meikle (2000) for SNe Ia. \\
The variation of NIR colors roughly follows a characteristic path in the $J-H$ 
vs.\ $H-K$ diagram, as can be seen in Fig.~\ref{f:nir-col-col}. 
Figure~\ref{f:nir-col-comp} compares NIR color-color diagrams, after 
dereddening, for 6 SNe Ia which are well sampled at $JHK$ (from Meikle 2000, 
Candia et al.\ 2003,
Krisciunas et al.\ 2003).  All resemble SN~2000E in their shift of colors
from below to above the blackbody locus.
We identify a common evolutionary sequence in two phases. 
First, earlier than $B_{\rm max}$, the data points tend to lie below the main
sequence locus and move towards bluer $J-H$ and redder $H-K$. 
It seems that the path followed by SN2000E in the color-color diagram well 
before maximum light  can be extrapolated backward in time up to the 
blackbody locus at a color temperature of 10000--20000 K. This is what is 
obtained in Fig.~\ref{f:nir-col-col} and Fig.~\ref{f:nir-col-comp} through the 
intersection of the straight line connecting the data points earlier than 
phase $ -5 $d, and the blackbody locus. The same is suggested by the NIR 
colors of a few of the SNe that have been sampled at very early epochs and may 
reflect the resemblance of spectra longward of $ B$ to blackbody curves of $ > 
10000$ K (see Sect.~\ref{phot:bm}).
Soon before the earlier NIR maximum, the data points cross the 
main sequence locus and move above it.
The change occurs roughly perpendicular both to the reddening law and to the 
blackbody sequence. However, it is obvious that this picture differs for each 
SN in some details. For example the maximum displayed color excess before 
$B_{\rm max}$: a slow decliner like SN~2000E shows the largest color excess 
during the first phase, whereas a fast decliner
like SN~1986G shows almost no color variations before $B_{\rm max}$. 
From these data it
is difficult to assess whether some of the differences are related to the 
declining rate. Past the secondary NIR peak, the differences in color 
evolution become even more noticeable probably because of the beginning of 
the nebular phase and the growing importance of line emission.
Anyway, some of the differences as well as part of the displayed ``oscillations'' could
be due to errors and color effects on the photometry. These problems should be 
considered in more detail.

%
%

%
%

%
%

\subsection{Reddening}
\label{Red}
From the maps of Schlegel, Finkbeiner, \& Davis (1998) we obtain a
Galactic extinction $A_{B} = 1.57$ mag towards NGC~6951. As noted in
Di Carlo et al.\ (2002), these maps appear to systematically overestimate
the extinction where $E(B-V) \ge 0.15$ mag, which is also the case for 
NGC~6951. The value from Burnstein \& Heiles (1982), $A_{B} = 0.88$ mag, is 
more consistent with the equivalent width of the interstellar NaID line in the 
spectra of SN~2000E (which we 
present in Fig.~\ref{spec:fig}). NaID lines are also found at the systemic 
velocity, suggesting absorption from the host galaxy, with EW(NaI) $\sim 
0.6$ \AA, which, following the relations given in Turatto, Benetti \& 
Cappellaro 2003, yields $0.1 < E(B-V) < 0.4$ mag. Using the extinction law 
from Cardelli, Clayton \& Mathis  (1989), assuming $R_{V}=3.1$, the total $E(B-V)$ is 
$\sim 0.32-0.62$ mag ($\sim 0.5-0.8$ mag using the value from 
Schlegel et al.\ 1998).

Purely photometric methods agree with the spectroscopic constraints.
Following the prescriptions by Phillips et al.\ (1999), we determine
the reddening for \e. This analysis yields:
$E(B-V)_{\rm max} = 0.65 \pm 0.06$ mag
[$E(B-V)_{\rm max} = 0.55 \pm 0.06$ mag for the sparser CT data set; see
Sect.~\ref{odr}];
$E(V-I)_{\rm max} = 0.73 \pm 0.12$ mag [i.\ e. $E(B-V)=0.45$ mag assuming 
$R_{V} = 3.1$]. 
Although we have only one $B-V$ point that can be shifted
on the Lira relation [yielding $E(B-V)_{\rm tail} = 0.63$ mag], just at the 
beginning
of the time interval defining the locus, we can circumvent this
by using a template $B-V$, that is fit by $E(B-V)_{\rm tail} = 0.59$ mag. 

As for the $V$ minus NIR colors,
their fits to the relations found by Candia et al.\ (2003) for slow declining 
Type Ia SNe
yield $E(V-J) = 1.08 \pm 0.02$, $E(V-H)=1.21 \pm 0.02$ and $E(V-K) = 1.32 \pm 
0.02$ mag. 
Assuming $R_{V} = 3.1$
and the reddening law of Cardelli et al.\ (1989), we find $A_{V}$ in the range
1.47--1.50 mag [$E(B-V) = 0.47-0.49$ mag].  
So also these values are consistent within the errors with
the spectroscopic indications. 
Using a lower $R_{V}$, e.\ g., $R_{V}=2$,
would result in $A_{V} \sim  1.4$ mag and $E(B-V) \sim 0.7$ mag, which lies above
the highest (and less likely) end of the spectroscopic range. 

In conclusion, different methods appear to constrain the reddening within a 
range $E(B-V) \sim 0.3-0.8$ mag.
All our determinations, including those discussed in 
Sect.~\ref{mlcs:sec}, are summarized in Table~\ref{red:tab}. In the
following, we will assume the average as the most likely value, yielding
$A_{V} = 1.55$ mag for $R_{V}=3.1$.

\subsection{MLCS fits and estimates of distance}
\label{mlcs:sec}
The distance to NGC~6951 is discussed by Vink\'o et al. (2001). They
report Tully-Fisher determinations 
of the distance modulus $\mu_{0}=31.85-31.91$ mag from the literature; based on their own data
of SN~2000E, they revise the distance modulus to $\mu_{0}=32.59 \pm 0.5$ mag. 
Herein, we attempt to refine the estimates of extinction and distance using 
our own data set of SN~2000E and different techniques. 

The MLCS
method is defined by Riess et al.\ (1996). Assuming the reddening law given
by Cardelli et al.\ (1989) for $R_{V}=3.1$ and using the linear solution
given by Riess et al.\ (1996), we performed 3 different fits, one to the TNT
data set only, one shifting $B$ by 0.1 mag from the TNT onto the CT values
and one shifting $B$ by 0.1 mag from the CT onto the TNT values. We obtained
a $\chi^{2}$ ranging from 173 to 727, but these large numbers are obviously 
due to the smallness
of some of the photometric errors. So we repeated the 3 fits increasing the
minimum error to 0.05 mag; this yielded $\chi^{2}=34-64$ (reduced $\chi^{2} \sim 1$).
Summarizing the six results by using the average values
and their standard deviations, we found $A_{V} = 1.42 \pm 0.07$ mag
[i.\ e., $E(B-V)=0.46$ mag],
$\mu_{0} = 32.11 \pm 0.05$ mag and $\Delta=-0.30 \pm 0.08$ (the latter is the 
$V$-difference at
maximum between SN~2000E and a ``fiducial'' Type Ia SN). Note that the formal 
errors on $\mu_{0}$ from the fits ($\sim 0.2$ mag) are always larger than the 
standard deviation of the six values.
We also performed the fits varying $R_{V}$ from 2 to 5 and checked that smaller values
of $R_{V}$ give a slightly smaller $\chi^{2}$. E.\ g., using only the TNT data, the 
smallest $\chi^{2}$
is found for $R_{V}=2.0$, with $A_{V} = 1.09 \pm 0.04$ mag [i.\ e., 
$E(B-V)=0.55$ mag],
$\mu_0 = 32.51 \pm 0.13$ mag and $\Delta=-0.38$. However, the difference in 
$\chi^{2}$ between $R_{V}=2.0$ and 3.1 is only $\sim 0.6$ \%
($\sim 1$), and we managed to mimick
a decrease of $R_{V}$ in test fits with templates 
simply using slightly greater epochs of maximum 
at $B$ than the actual one. These also produced slightly smaller 
$A_{V}$'s and slightly greater distance moduli. 
It is noteworthy that the MLCS method yields extinction values 
in agreement with the estimates given in Sect.~\ref{Red} and most of the estimated 
distances lie between 32.1--32.2 mag.

\begin{figure}
\includegraphics[height=12cm,angle=270]{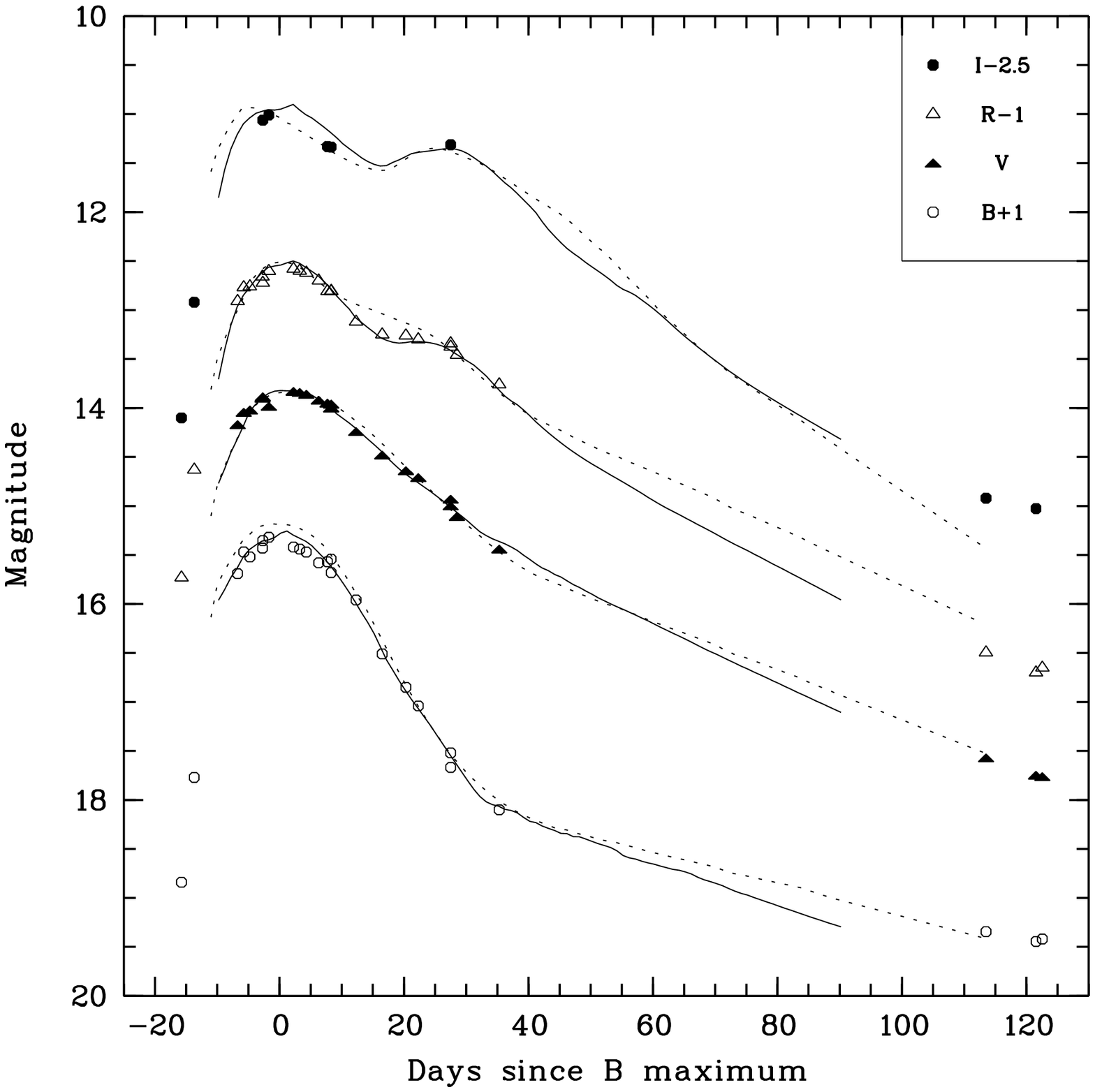}
\caption{$BVRI$ light curves of SN~2000E overlaid with the fitting light 
curves (solid line) from the updated MLCS method [$\Delta=-0.54 \pm 0.03$, 
$\mu_{0} = 32.42 \pm 0.06$ mag, $E(B-V) =0.45 \pm 0.03$ mag]. The light curves 
from one of the MLCS fits (dotted line) is also drawn [$\Delta=-0.19$, 
$\mu_{0} = 32.05 \pm 0.14$ mag, $A_{V} = 1.34 \pm 0.06$ mag]. 
\label{mlcs:f}}
\end{figure}

The linear MLCS method uses training vectors based on a very limited number
of SNe. An updated version employing a larger data set
is described in Riess et al.\ (1998). 
We performed fits to the $BVRI$ light curves following the
updated MLCS method, with the 
templates kindly provided us by Adam Riess. Each model light curve includes 
also a second order term in $\Delta$ (see Eq.~A6 in Riess et al.\ 1998) and 
we adopted the extinction law
computed by Nugent, Kim \& Perlmutter (2002) for Type Ia SNe as a function 
of epoch. 
We used a routine which simultaneously fits four model light curves to the four 
optical 
ones determining the four free parameters $\Delta$, $\mu_{0}$, $E(B-V)$ and 
$t_{\rm max}$
through minimization of the deviations between data and model weighted by the 
sum
of the squared photometric errors and the two-point correlation terms. The zero points
of the templates have been set invoking $B=-19.45$, $V=-19.45$, $R=-19.50$ and
$I=-19.10$ mag at $t_{\rm max}$ (the epoch of maximum light at $B$). We repeated
the calculations varying the data set
in the 6 ways already selected for the linear MLCS, always obtaining
quite similar results, but with large $\chi^{2}$ values ($\sim 3000$) when
the photometric errors are unchanged and acceptable ones ($\chi^{2} < 105$, i.\ e.,
reduced $\chi^{2} < 1.7$) when the minimum errors are increased to 0.05 mag.
Still using average values and standard deviations to summarize
the six sets of results, we obtained $\Delta=-0.54 \pm 0.03$, $\mu_{0} = 32.42 
\pm 0.06$ mag, $E(B-V)
=0.45 \pm 0.03$ mag and $t_{\rm max}=577.2 \pm 0.2$. The standard deviations are
always of the same order (or slightly greater) than the formal fit errors. 
Hence, also the updated MLCS method yields an $E(B-V)$ which agrees with the estimates
given in Sect.~\ref{Red}, but the SN appears even more overluminous (more 
distant) 
with respect to the results of the linear MLCS method. 
The same trend is obtained by
Vink\'o et al. (2001), although they find an {\em underluminous} object through the linear
MLCS method. The template LCs at $BVRI$ obtained by the updated MLCS fit appear
to reproduce the shape of the observed LCs, as can be seen in
Fig.~\ref{mlcs:f}, better than those from the linear MLCS fit.
 
\begin{deluxetable}{cc}
\tablecaption{Reddening as determined from the different methods used.
  All assume $R_{V}=3.1$.
         \label{red:tab}}
	 \tablewidth{0pt}
	 \tablehead{
	 \colhead{Method} & \colhead{$E(B-V)$} \\
	 \colhead{ } & \colhead{(mag)} \\
	 }
	 \startdata
NaID lines & $0.32-0.62$ \\
$E(B-V)_{\rm max}$ & $0.55-0.65$ \\
$E(V-I)_{\rm max}$ & $0.45$ \\
Lira & $0.59-0.63$ \\
$E(V-J)$ & $0.49$ \\
$E(V-H)$ & $0.48$ \\
$E(V-K)$ & $0.48$ \\
MLCS & $0.46$ \\
updated MLCS & $0.45$ \\
max.\ $BVI$\tablenotemark{a} & $0.42 - 0.45$ \\
max.\ $BVI$\tablenotemark{b} & $0.45 - 0.48$ \\
\hline
average value & $0.50 \pm 0.09$ \\
\enddata
\tablenotetext{a}{Fit to max.\ $BVI$ (Hamuy et al.\ 1996a).}
\tablenotetext{b}{Fit to max.\ $BVI$ (Phillips et al.\ 1999).}
\end{deluxetable}

%
%

Further
constraints can be put on the previous results by using other available 
techniques for the distance estimate. Hamuy et al.\ (1996a) and
Phillips et al.\ (1999) find updated relations between absolute
$BVI$ magnitudes at maximum and $\Delta m_{15}(B)$, so 
$A_{V}$ and $\mu_{0}$ can be simultaneously estimated 
by fitting $M_{\lambda} + (A_{\lambda}/A_{V})A_{V}+
\mu_{0}$ to the $BVI$ magnitudes given in Tab.~\ref{peak_date}.
Still, we adopted the reddening law by Cardelli et al.\ (1989) assuming
$R_{V}=3.1$. From Hamuy et al.\ (1996a) we obtain $A_{V} =
1.39$ mag and $\mu_{0} = 31.88$ mag ($A_{V} = 1.30$ mag and $\mu_{0} =
31.94$ mag for $B=14.30$ mag), whereas the Phillips et al.'s (1999) yields
$A_{V} = 1.49$ mag and $\mu_{0} = 32.08$ mag ($A_{V} = 1.39$ mag and
$\mu_{0} = 32.14$ mag for $B_{\rm max}=14.30$ mag). A conservative estimate of
the formal uncertainty for $\mu_{0}$ is $\sim 0.2$ mag. Note that
Phillips et al.\ (1999) give only the variation of absolute magnitude
with $\Delta m_{15}(B)$ for ones where 
$\Delta m_{15}(B) = 1.1$ mag (however quoted by Stritzinger et al.\ 2002).
Our estimated extinction agrees well with that from spectroscopy,
which confirms the reliability of the results. 

Checks can also be performed using our NIR data. Meikle (2000) finds
that 6 out of 8 examined Type Ia SNe have similar coeval
absolute magnitudes at $JHK$ ($13.75$ days past $B_{\rm max}$). Using
the mean values and our interpolated magnitudes for that epoch, we
obtain $m - M = 31.90, 32.23$ and $32.29$ mag at $J$, $H$ and $K$
respectively. Note that $J$ is discrepant with respect to $H$ and $K$
for any reddening law, in
the sense that no common $A_{V}$ exists yielding a single 
distance modulus. Using our preferred value of $A_{V} = 1.55$ mag results in
$\mu_{0} = 31.46 \pm 0.12, 31.94 \pm 0.11$ and $32.11 \pm 0.11$ mag from 
$J$, $H$ and $K$
respectively. 
Although $H$ and $K$ estimates agree with previous determinations,
$J$ produces a much smaller value. This suggests that SN~2000E must have a 
brighter  intrinsic $J$ magnitude with respect to the sample of Meikle (2000), 
if the distance modulus derived from $J$ needs to be larger (so as to agree 
with those derived from H and K).
Clearly, more data are 
needed to establish meaningful absolute magnitudes at NIR wavelengths. 
Furthermore, whereas the previous distance determinations are based on the 
Cepheid scale as given by Sandage et al.\ (1996) and Saha et al.\ (1997), 
Meikle (2000) adopts the revised Cepheid scale of Gibson et
al.\ (2000), whose zero point is $0.12 \pm 0.07$ mag less (i.\ e., the 
$\mu_{0}$ values according with the older scale are to be decreased by 
0.12 mag).
 
Similarly,
Krisciunas et al.\ (2003) show that the absolute magnitude 
of SNe Ia at $H$-band 10 days past maximum $B$ light is a flat function of
$\Delta m_{15}(B)$ and find a mean value $H = -17.91 \pm 0.05$. Assuming $A_{V}
= 1.55$ mag and the reddening law of Cardelli et al.\ (1989), 
along with an interpolation of our $H$ data to this epoch, this
yields $\mu_{0} = 31.68 \pm 0.11$ mag. This is significantly less than all 
other determinations, except for the one from the absolute magnitude at $J$. 
However, the discrepancy appears to be due to the different Cepheid
scale used by the authors. In fact, the given $H$
is on the Cepheid scale as revised by Freedman et al.\ (2001);
a comparison between their distances (see their Tab.~3) and those listed
by Sandage et al.\ (1996) and Saha et al.\ (1997) and adopted in the
previous methods indicates a mean difference of $\sim 0.4$ mag, in the sense
that distance moduli from Freedman et al.\ (2001) are smaller.

In summary, different techniques produce distance moduli ranging
from $\sim 31.5$ to $\sim 32.4$ mag, with most of them consistent
with $\sim 32.0$ mag. When the calculation of $A_{V}$ is also 
involved, the estimated extinction is always consistent with the constraints
given in Sect.~\ref{Red}. This appears to lie within $\sim 1.3 - 1.5$
mag, so its uncertainty cannot contribute more than $\sim 0.2$ mag to
the estimated variations of $\mu_{0}$.\\
An average of all values found through 
the brightness and colors of SN~2000E yields $\mu_{0} = 32.0  \pm
0.3$ mag. 
%
%
However, using the determinations consistent with the older Cepheid scale
of Sandage et al.\ (1996) and Saha et al.\ (1997) yields $\mu_{0} = 32.14  
\pm 0.21$ mag. This choice is being made because our estimates are mainly based
on the MLCS method, whose results are being so far reported in the literature
mostly with respect to the older scale. Anyway, MLCS distances can be
shifted to the latest Cepheid scale by subtracting $\sim 0.2$ mag from
$\mu _{0}$ (A.\ Riess, private communication).

\subsection{Bolometric behavior of SN~2000E}
\label{bolo:sec}

All measurements in the $UBVRIJHK$ bands were dereddened according to
the extinction $A_{V} = 1.55$ mag derived in Sect.~\ref{Red}. The optical 
magnitudes were converted to fluxes according to Bessel (1979) and,
in the NIR bands, the calibration given for the UKIRT standard. 
We constructed the curve of the bolometric luminosity using
the results of spline fits to the LCs in order to have homogeneous
sets of magnitudes at all wavelengths for each epoch.
$U$ was estimated by assuming $U-B=0.2$ mag before JD+575 (according to
the earliest measurement) and by approximating the LC using three different
segments through the available data points after JD+575.

\begin{figure}
\epsscale{0.8}
\plotone{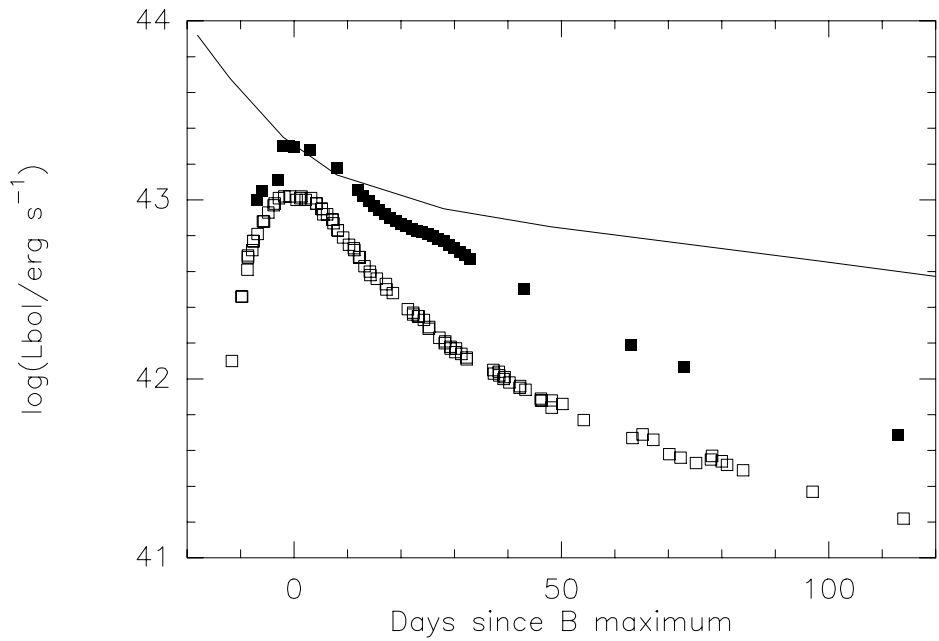}
\caption{Bolometric luminosity of SN~2000E (filled squares) as derived by flux
(uvoir) integration. For the sake of comparison, the bolometric light curve of 
the fast declining SN~1994D is shown (open squares; Salvo et al.\ 2001).
The radioactive decay luminosity (Branch 1992) for 0.9 M$_{\sun}$
of $^{56}$Ni and a rise time of 18 days is also drawn.
\label{bolo:fig}}
\end{figure}

Experiments with blackbody fits according to the technique described
in Di Carlo et al.\ (2002) indicates that the SN emission is reminiscent of a  
blackbody only at earlier epochs (before $B_{\rm max}$). So, we determined the 
bolometric luminosity just as the area of the trapezium connecting the 
$UBVRIJHK$ fluxes. 
We used a distance of $26.8 \pm 2.6$ Mpc, as previously estimated. 
We did not use $U$ magnitudes before JD+575; after this epoch, the same values
as estimated for the blackbody fits were adopted  and we set the flux to 0 at
3000 \AA\ (see Suntzeff 1996). 
Table~\ref{bolo:tab} lists the (uvoir) bolometric luminosity and 
Fig.~\ref{bolo:fig} shows the time evolution of the uvoir luminosity, compared 
with that of SN~1994D, a fast declining SN Ia (Salvo et al.\ 2001).
Integrating the bolometric curve from JD+561 to JD+690 (i.\ e., 16 days
before $B_{\rm max}$ to 113 days past it) yields $3.6 \times 10^{49}$ ergs.
%
%
\begin{deluxetable}{ccccc}
\tablecaption{Bolometric (uvoir) luminosities of SN~2000E.  
\label{bolo:tab}}
	 \tablewidth{0pt}
	 \tablehead{
	 \colhead{Julian Day} & \colhead{Epoch}  & \colhead{Bolometric}\\
	 \colhead{ } & \colhead{ } & \colhead{Luminosity} \\
	 \colhead{ } & \colhead{ } &  \colhead{(uvoir)} \\
	 \colhead{(2451000+)}& \colhead{(days)} &\colhead{($10^{43}$ erg s$^{-1}$)}
	 }
	 \startdata
$570$ & $-7$  & $1.004$ \\
$571$ & $-6$  & $1.122$ \\ 
$574$ & $-3$  & $1.294$ \\ 
$575$ & $-2$  & $1.998$ \\ 
$576$ & $-1$  & $1.988$ \\
$577$ &$ 0$   & $1.985$ \\ 
$580$ &$ 3 $  & $1.893$ \\ 
$585$ &$ 8$   & $1.516$ \\ 
$589$ &$ 12 $ & $1.133$ \\ 
$604$ &$ 27$  & $0.603$ \\ 
$620$ &$ 43$  & $0.316$ \\
$640$ &$ 63$  & $0.156$ \\ 
$650$ &$ 73 $ & $0.116$ \\ 
$690$ &$ 113$ & $0.049$ \\
\enddata
\end{deluxetable}

The uncertainty, excluding that inherent to distance and 
reddening, should be below $\sim 10$ \%. 
Nevertheless, before JD+575 (phase $-2 $d) uvoir fluxes should be corrected of 
a factor up to $\sim 1.2$ due to the lack of measurements at $U$.  This 
explains the small break in the uvoir curve around phase $-2 $d  evident in 
Fig.~\ref{bolo:fig}. 
Leibundgut \& Pinto (1992) note that the use of broad-band filters to 
integrate the uvoir fluxes tends to overestimate the bolometric luminosity 
for emission-line objects; they infer, however, that the effect should be 
$\sim 10$ \%. At late time (beyond $\sim 35$ days after $B_{\rm max}$), the 
derived bolometric luminosities are less reliable, due to the poorer 
representation of the true light curves given by the spline fits between 
JD+612 and JD+690.
The estimate of the total initial mass of $^{56}$Ni follows from the 
prescription  
of Arnett (1982), according to which the bolometric luminosity at maximum 
equals the
radioactive input of $^{56}$Ni. However, this requires a knowledge of the 
rise time; 
we assumed it to be 18 days, which is the least value consistent with the
earliest detection of SN2000E. Using the relation given by Branch (1992) for 
the radioactive decay of $^{56}$Ni, we obtained a $^{56}$Ni mass of $\sim 0.9 
\pm 0.2$ M$_{\sun}$. 
The given uncertainty only reflects that in $\mu_{0}$ and the lower value of
$^{56}$Ni mass would be obtained from the Tully-Fisher distance. Since
the uvoir integration is performed from 3000 \AA\ to $2.2$ $\mu$m, it is very 
unlikely that the bolometric flux (hence, the $^{56}$Ni mass) is 
underestimated by more than 10 \% at maximum light. Increasing the rise time 
to 21.4 days (the value envisaged by Riess et al.\ 1999 for $\Delta  
m_{15}(B)=0.94$) would result in a $^{56}$Ni mass 19 \% larger. Another error 
of $< 30$ \% accounts for an uncertainty of 0.3 mag in $A_{V}$.
The radioactive decay curve (Branch 1992) for 0.9 M$_{\sun}$ of $^{56}$Ni 
assuming a rise time of 18 days is shown in Fig.~\ref{bolo:fig}. 

%
%

\subsection{Spectroscopy} 
\label{spectroscopy}

\begin{figure}
\includegraphics[width=15cm,height=15cm,angle=0]{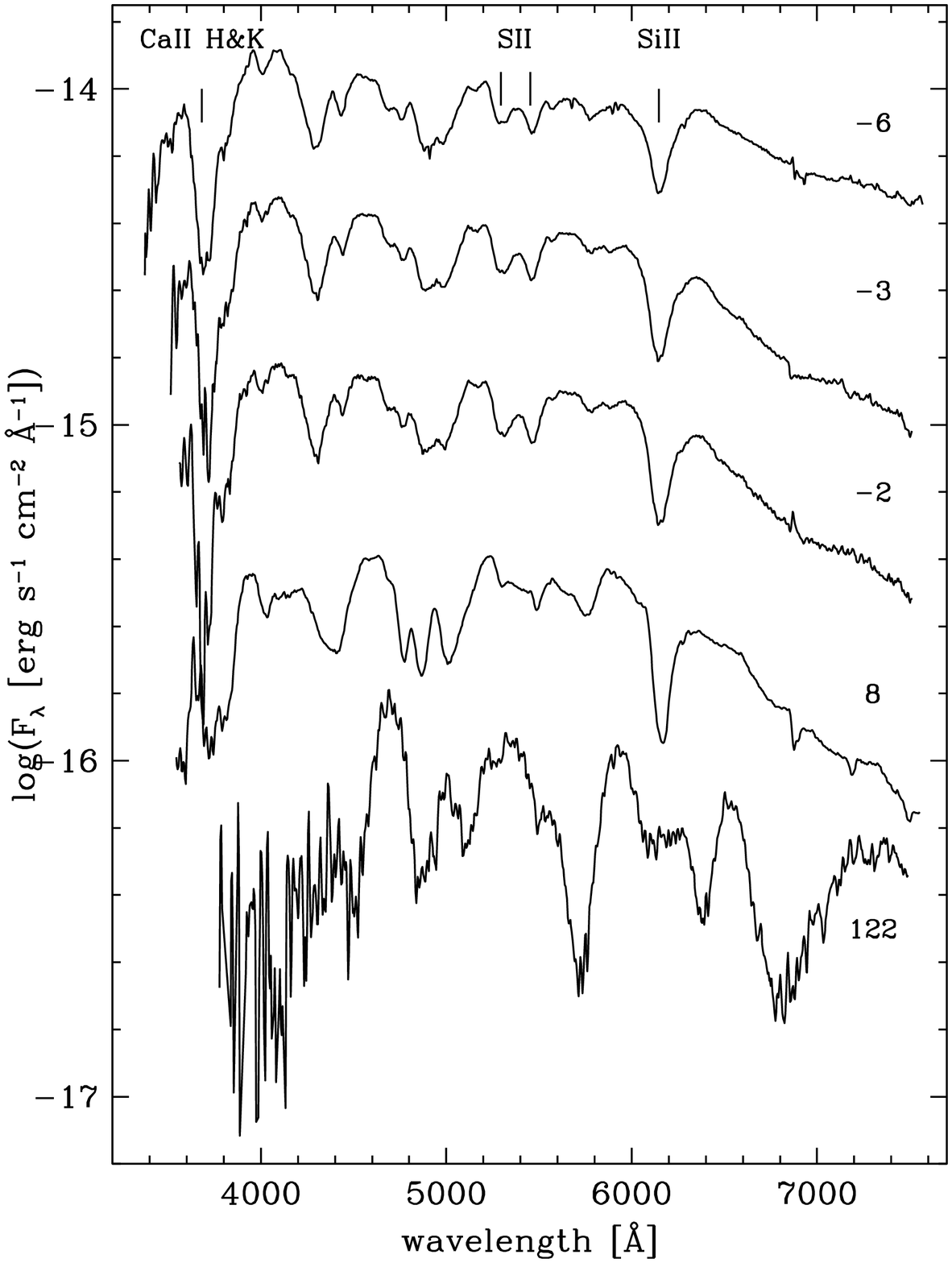}
\caption{
Spectral evolution of SN~2000E. Wavelength is in the observer
frame. The ordinate refers to the first spectrum, and the others have
been down-shifted by: 0.5, 1.0, 1.5 and 0.65, respectively. All the
spectra have been obtained with the Asiago~1.82m telescope +
AFOSC. The resolution is 25\AA~ as measured from the FWHM of the
night-sky lines. The flux calibration of the spectra has been checked
against the photometry and in the case of discrepancies the spectra
were adjusted. The lines discussed in the text are labeled.
\label{f:spec_evol}}
\end{figure}

Figure~\ref{f:spec_evol} illustrates the spectroscopic evolution of
SN~2000E from phase $-6d $to $+122 $d (i.\ e., time since $B_{\rm max}$).
%
%
The spectral evolution of SN~2000E follows the normal SNe Ia pattern
(cfr. Figure 18.3 of Wheeler \& Benetti 2000).
The SiII 6355 \AA\ and the CaII H\&K absorptions are always the most
intense features of the spectrum. Also typical is the evolution of 
the SII doublet at about 5000 \AA (labeled in Fig.~\ref{f:spec_evol}),
which remains
almost unchanged before maximum, while it is disappearing by phase$ +8 $d.
The photospheric expansion velocities deduced from the SiII~6355 \AA\ minima 
show a small decline from 10900 km s$^{-1}$ roughly 6 days before
maximum to 10300 km s$^{-1}$ at phase of $+8 $d.
Most SNe show changes in expansion velocities from -6 to +8 days in the range
$\sim 1000 - 2000$ km s$^{-1}$ (see, e.\ g., Fig.~10 of Patat et al.\ 1996 
and Fig.~11 of Salvo et al.\ 2001), whereas SN~2000E shows a decrease of only 
600 km s$^{-1}$. Therefore one may wonder whether this exceptional behavior 
of SN~2000E is manifest in other ways.

\begin{figure}
\includegraphics[width=8.5cm,]{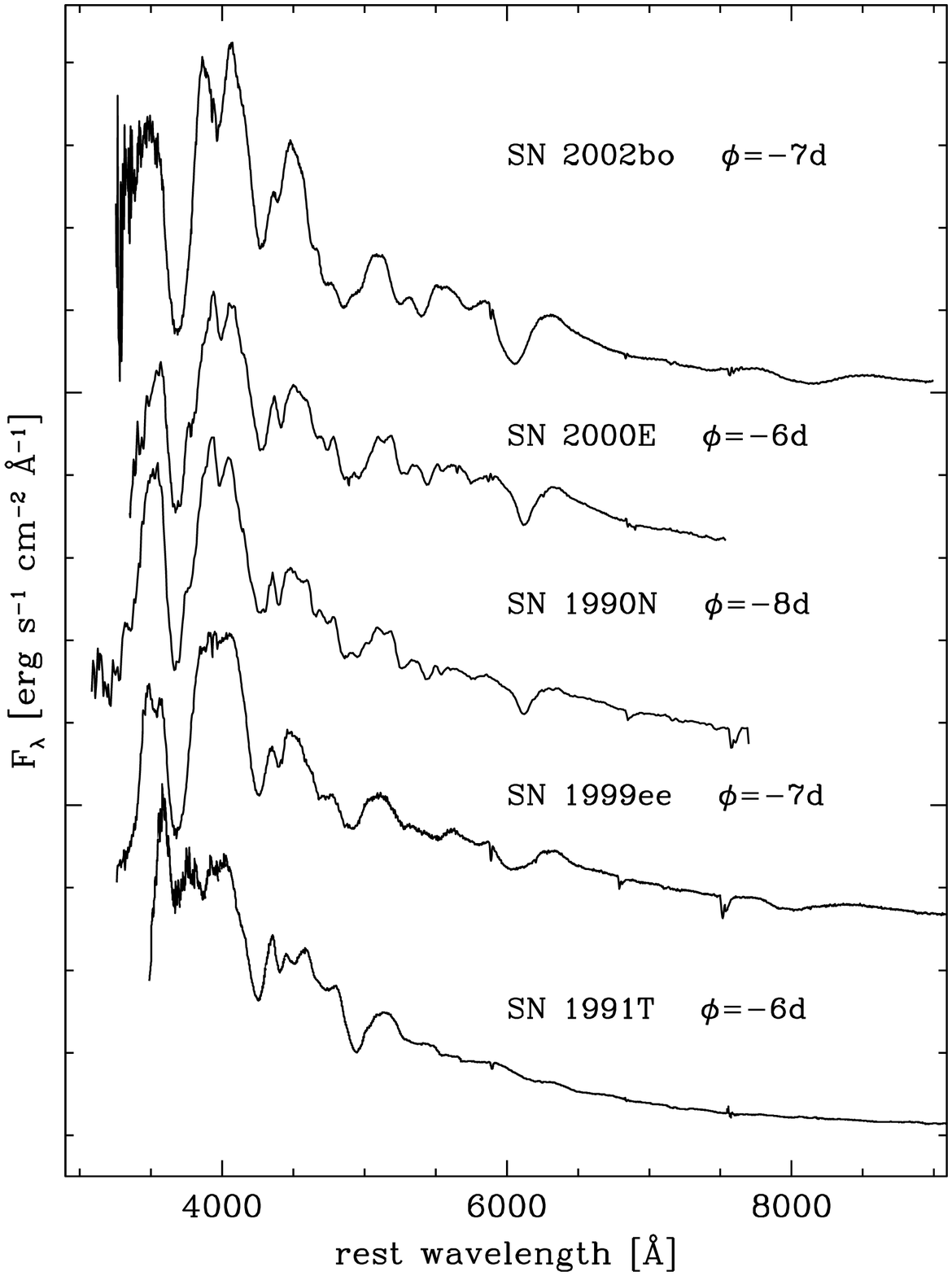}
%
%
\caption{Comparison between spectra of SNe Ia about a week before maximum
light. All the SNe shown have a similar $\Delta m_{15}$(B)$\sim
0.94-1.07$.  The data sources are: S. Benetti et al., in preparation
(SN~2002bo); Leibundgut et al. 1991 (SN 1990N); Hamuy et al. 2002
(SN~1999ee) and Mazzali et al. 1995 (SN~1991T). The spectra have been
dereddened and the wavelengths have been reported to the parent galaxy
rest frame.
\label{spec:fig}}
\end{figure}

Small differences between spectroscopically normal Type Ia SNe are seen as 
small spectral details when spectra at similar epoch are compared (Fig.~
\ref{spec:fig}). This is particularly true well before maximum (S. Benetti 
et al., in preparation).

Vink\'o et al. (2001) discuss spectral characteristics of overluminous
SN~1991T-type objects and their analogy with SN~2000E. This subgroup of Type
Ia SNe show peculiar pre-maximum spectra consisting of less prominent features 
with a few ionized Fe lines but without Si or S lines (Li et al.\ 2001).  
Conversely,
these lines can be easily identified in our pre-maximum (6 days before $B_{\rm
max}$) spectrum of SN~2000E taken with the CT. This is shown in
Fig.~\ref{spec:fig} along with others of comparable phase from three SNe Ia
(including SN~1991T) exhibiting the same $\Delta m_{15}(B)$. The spectra are
all in the rest frame of the host galaxies and have been dereddened [see
S. Benetti et al., in preparation, for SN~2002bo, Hamuy et al.\ 2003 for 
SN~1999ee, Mazzali, Danziger \& Turatto 1995 for SN~1991T although this has
been corrected for $E(B-V)=0.17$ mag following Phillips et al.\ 1999].
In the Figure, the spectra have been plotted following the intermediate 
mass elements (mostly CaII, SII and SiII) P-Cygni line strength: with 
SN~2002bo having the strongest lines,  down to the
peculiar SN~1991T with no intermediate mass elements lines in the
spectrum.
While all the SNe shown in the figure have similar continua, which
means they have similar photospheric temperatures, the SiII 6355 \AA\
trough shows different shapes: broad and slightly asymmetric in
SN~2002bo; with a composite profile (broad wing and narrow core) in
SN~2000E and SN~1990N; again, broad and slightly asymmetric in
SN~1999ee (even if it is less deep than in SN~2002bo); absent in
SN~1991T.
However, Figure \ref{spec:fig} shows that the spectrum of SN~2000E is
almost identical to that of the 
SN~1990N, that is a spectroscopically normal SN Ia 
with  $\Delta m_{15}(B) = 1.03$ mag (Lira et al.\ 1998). 
In fact, Hamuy et al.\ (2003) show
that not all slow-decliners are spectroscopically peculiar before $B_{\rm max}$
and distinguish between 1991T-like events and 1999aa-like events.

\subsection{The photosphere before maximum light}
\label{phot:bm}
The available spectroscopic data for SNe Ia indicate that longward
of $B$, pre-maximum spectra are roughly approximated by blackbody
spectra, although the fit quality worsens with time (see Meikle et al.\
1996, Hamuy et al.\ 2003; also Suntzeff 2003). 
Pre-maximum NIR spectra of SN~1999ee, 
a slow-declining SN Ia, are remarkably almost featureless (Hamuy et 
al.\ 2003). But, also in the optical, pre-maximum spectra 
down to $B$-band wavelengths appear much less line-dominated than 
those past maximum, as can be seen from those shown in 
Fig.~\ref{f:spec_evol} and \ref{spec:fig}.
Figure~6 of Meikle et al.\ (1996) suggests that the blackbody 
approximation is no longer satisfactory later than one week before 
$B_{\rm max}$, but it is acceptable probably until the
diffusion time remains longer than the elapsed one (see also Arnett 1982).
If this is true, a color temperature may be derived from a blackbody fit to 
the $BVRIJHK$ fluxes. In turn, this can be used in order to estimate the 
surface flux and the evolution of the photosphere during the rise time. After
 that epoch, the spectra shown in Meikle et al.\ (1996) and Hamuy et al.\ 
(2003) are evidently poorly fitted by blackbodies. 
 
\begin{figure}
\epsscale{0.4}
\plotone{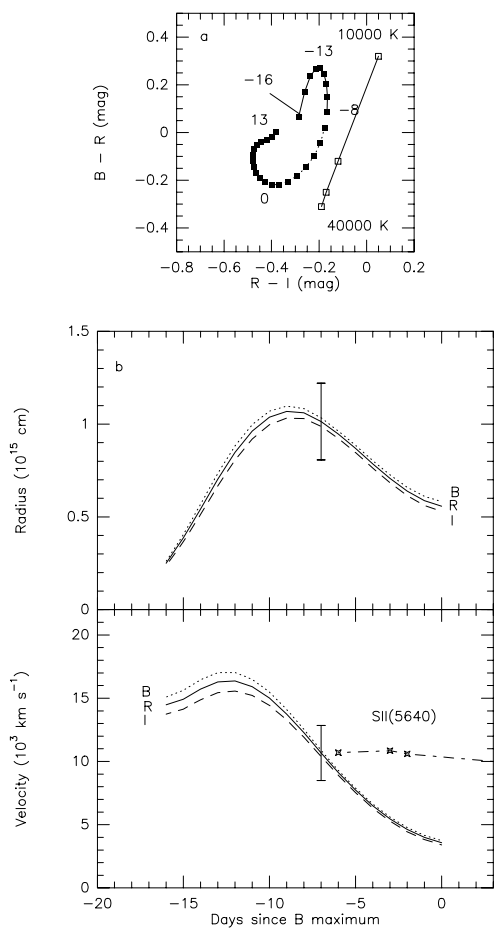}
\caption{{\bf (a)} Color-color ($B-R$ vs.\ $R-I$) diagram
 showing the time evolution of SN~2000E from phase $-16 $to
 $+13 $(in days; labeled); the blackbody locus for temperatures from
 10000 K to 40000 K is drawn (dotted line) for the sake of comparison.
 {\bf (b)} Photospheric radius (top) and velocity (bottom) from blackbody
 fits to $VRI$ only.
 Typical errorbars are shown at phase $=-7 $d.  The values have been
 calculated for each band by scaling the blackbody to observed
 flux ratio according to distance and reddening. The velocities
 obtained from the absorption of the SII line at 5640 \AA 
are also drawn (dash-dotted line).
\label{phot:fig}}
\end{figure}

Unfortunately, we do not have a full coverage before $B_{\rm max}$ at all
wavelengths, so we adopted the magnitudes from the spline fits to $BRI$ only. 
These were dereddened assuming $A_{\rm V} = 1.55$ mag.
We used the fitting procedure described in Di Carlo et al.\ (2002) in order
to derive the blackbody temperature; with three points only, it is more or less
equivalent to determine the slope of the SED from $B$ to $I$.
Once a temperature is derived as a function of phase, photospheric radii
are computed for each passband effective wavelength 
from the ratio of blackbody to observed (dereddened) flux scaled to the 
assumed distance. The velocity is then obtained assuming a rise time of 18 
days. These are illustrated in Fig.~\ref{phot:fig}b.
The photospheric velocities are of the same order of magnitude
as the expected ones 
($10000-20000$ km s$^{-1}$). However, a comparison with the velocities 
obtained from the absorption of the SII line at 5640 \AA\ (also drawn in 
Fig.~\ref{phot:fig}b) in the spectra of SN~2000E 
shows that the two estimates are similar 6 days before $B_{\rm max}$, but 
whereas the spectroscopic values decline quite slowly at least until two 
days after $B_{\rm max}$, the photometric values fastly decrease. This may be 
due to the increasing departure of the SED from a blackbody when approaching 
$B_{\rm max}$.
If taken at face value, our blackbody estimates would suggest
that the photosphere started receding not later than JD+569--JD+570  
(phase$ -8$ to$ -7 $d). 
This behavior is determined by the changing in temperature, first decreasing
from $\sim 16000$ K to $\sim 12000$ K between JD+561 (phase $-16 $d) and
JD+565 (phase $-12 $d), then increasing again.
This causes the reverse of velocity and, later, of radius evidenced
in Fig.~\ref{phot:fig}b. It is clear from the color-color diagram of
Fig.~\ref{phot:fig}a, after dereddening according $A_{V}=1.55$ mag and
$R_{V}$=3.1, that the temperature evolution is mainly driven
by the increase of $B-R$ from phase $-16$ to $-13 $d and the subsequent decrease
within the range of blackbody colors for $T=10000$ K to $40000$ K (also shown 
in figure). Although $R-I$, unlike $B-R$,  does not lie within this range, the 
difference is only 0.1-0.2 mag and the approximation may still be good.
The increasing departure of $R-I$ from the locus of blackbody colors starting
at phase $-6 $d and the decrease of $B-R$ towards values typical of
temperatures of $\sim 40000$ K may
be just other symptoms of the worsening of the blackbody approximation.

%
%

\section{Discussion}
\label{disc}

A careful comparison of optical and infrared LCs with templates and data
from other SNe Ia has shown that SN~2000E is what is usually referred to
as a slow decliner.
We can check the overluminous nature of SN~2000E through its absolute 
magnitudes at peak as derived from the only independent distance estimates 
available (based on the Tully-Fisher relation). These may be compared with 
those of an overluminous SN Ia on the same distance scale. SN~1991T is well 
suited to this; using the photometry from Lira et al.\ (1998) and the 
distance (which is also in the Tully-Fisher scale) and extinxtion given by 
Mazzali et al.\ (1995), we obtain 
$B_{\rm max}=-19.45$ mag, $V_{\rm max}=-19.55$ mag, $R_{\rm max}=-19.55$ mag 
and $I_{\rm max}=-19.15$ mag. Even with a reddening as low as $A_{V}=1.3$ mag, 
SN~2000E would have $B_{\rm max} \le -19.20$ mag, $V_{\rm max}=-19.40$ mag, 
$R_{\rm max}=-19.30$ mag and $I_{\rm max}=-19.00$ mag. Most of the values are 
only $\la 0.2$ mag larger, hence it is quite likely that SN~2000E is an 
overluminous Type Ia SN, in the sense that it lies at the high luminosity end
of the observational range.
 
It appears that the photometric behavior of SN~2000E is quite standard for a slow-declining
overluminous Type Ia SN.
The reduced extinction at NIR wavelengths may pave the way towards a
significant reduction of uncertainty in distance determinations using SNe Ia as
standard candles, if
relations between morphological features of LCs and absolute magnitudes do 
exist. The first results (e.\ g., Meikle 2000) are encouraging, but more work 
is needed, especially more NIR observations of a number of representative 
objects which allow one to understand the complex behavior of LCs around
the two peaks as a function of luminosity. Yet, the characteristic shape of 
light curves in the NIR appears potentially suitable for testing the physical 
mechanisms occurring in Type Ia SNe. These mechanisms should account not 
only for the double-peak morphology of LCs, but also for the color evolution 
at NIR wavelengths, such as the ``crossing'' of main sequence in a color-color 
diagram during the period spanning the first peak and the minimum between 
peaks. 
This appears as the main common behavior of Type Ia SNe in a $J-H$ vs.\ 
$H-K$ diagram (see Fig.~\ref{f:nir-col-comp}). NIR spectra (Meikle 
et al.\ 1996, Hamuy et al.\ 2003) indicate that the crossing is 
mainly caused by a 
flux deficit around $1.2$ $\mu$m which develops past $B_{\rm max}$.
In the case of SN~2000E,
not only the same double-peak morphology exhibited at $JHK$ is
apparent in the $R$ and $I$ LCs (see Fig.~\ref{f:LC}), 
but the same physical mechanism involved does
appear to affect the light curve at $V$, as well, and this is clearly noticeable in the 
$B-V$ colors of SN~2000E (Fig.~\ref{f:col}). In fact, 
from the time of the first NIR peak to 
that of the NIR minimum, $B-V$ gets redder roughly linearly; then this 
reddening continues with a steeper slope, until the time of the secondary NIR 
peak, which appears to coincide with that of the peak at $B-V$. Afterwards, the
 Lira sequence starts.
This analogy between $V$ and $JHK$ must not be unexpected, in view of the 
existence of well defined loci in the $V$ minus NIR diagrams both for 
mid-decliners and for slow-decliners (see Sect.~\ref{Red}).

The effect imprinting all LCs longward of $V$ leaves its
signature also in the bolometric curve showing up as a well marked shoulder 
around the epoch of the secondary maximum (Fig.~\ref{bolo:fig}). The 
bump on bolometric LCs $\sim 25$ days past $B_{\rm max}$ was pointed out 
by Suntzeff (1996) and Contardo, Leibundgut \& Vacca (2000). Although the 
strength of this feature varies even among overluminous SNe Ia, it seems that 
underluminous SNe Ia do not exhibit it (see Fig.~5 of Contardo et al.\ 2000, 
also Suntzeff 2003). 
 All this 
suggests that {\em the double-peak structure is not a mere redistribution of 
photons from blue to red wavelengths}. As noted by Candia
et al.\ (2003), both the energy source for the bolometric LC and the optical 
depth to gamma rays are monotonically declining at the epoch of the shoulder 
(see also our Fig.~\ref{bolo:fig}), so it must be associated with cooling 
mechanisms and not to energy input.
Two explanations have been suggested for the NIR double-peak morphology. 
Hoflich, Khokhlov \& Wheeler  (1995) find that in some of their delayed detonation models 
the photospheric radius $r$ still increases well after $B_{\rm max}$. 
Since the NIR flux is roughly $\sim r^{2} \times T_{\rm eff}$,
the decrease in $T_{\rm eff}$ is at a point compensated by the increase of $r$.
Conversely, Pinto \& Eastman (2000b) propose that the photosphere recedes 
rapidly to the center of the supernova in the NIR. Hence, 
after peak the temperature of the photosphere and
the ionization state of the gas decrease leading to a reduction in opacity 
and to an increase in emissivity at longer wavelengths. This allows energy 
previously stored within the SN to rapidly escape in the infrared. This view 
seems to be confirmed by the most recent spectroscopic data (Hamuy et al.\ 
2003) and would explain the $J$ deficit causing the ``crossing'' in $J-H$ vs.\ 
$H-K$ diagrams. Moreover, if our finding of a receding photosphere well 
before $B_{\rm max}$ is real, this would also favor the suggestion of Pinto \& 
Eastman (2000b). 
 
The noticeable slowing
down of the SiII velocities in all SNe around -5 days is plausibly due to the
fact that the cores of that absorption line are being formed above the
photosphere and therefore at higher velocities than the photosphere, which in
fact has receded quickly consistent with the proposal of Pinto \& Eastman. 
It is interesting to note that the velocities of SN~1990N and SN~1999ee (see 
Fig.~2 of Hamuy et al.\ 2003), obtained from the absorption minimum of the 
SiII line at 6355 \AA, exhibit a strong decrease from $\sim 16000$ to $\sim 
10000$ km s$^{-1}$ until $\sim$5 days before $B_{\rm max}$. 
This early decrease is quite similar to that found for SN~2000E using 
the blackbody approximation. The same behavior is found also in other Sne Ia 
(Salvo et al.\ 2001, in particular their Fig.~11).
Whether a particular distribution of Si in the envelope could plausibly have
a similar effect is not clear. 
The reasons for the different behavior of the
velocities among SNe Ia is still unknown and remains to be elucidated. 
This is obviously a critical point which needs also 
a greater knowledge of the physics of SNe Ia envelopes.

SN~2000E appears to fit a phenomenological sequence describing the behavior of
SNe Ia, which may be summarized as follows. The luminosity is linked
to the mass of synthesized $^{56}$Ni (Arnett's Rule). Overluminous objects show
slow declining light curves at $B$, whereas the opposite occurs for 
underluminous objects. Increasing the luminosity, the double-peaked morphology 
of NIR LCs develops and the secondary maximum tends to occur later (Phillips 
et al.\ 2003); it also extends shortward, producing a secondary peak at $I$ 
and a shoulder at $V$. At the same time, this enhances an inflection in the 
bolometric LC, as well.
Whether the sequence has a spectroscopic analogue is still debated. 
Hamuy et al.\ (2003) conclude that the photometric properties of luminous SNe 
Ia cannot be used to predict spectroscopic peculiarities.
A number of physical processes at work may account for variations of this 
picture. For example, the kinetic energy may not be strongly coupled to the 
mass of $^{56}$Ni; whereas $^{56}$Ni drives the luminosity, the kinetic energy 
may affect the cooling of the photosphere and, hence, the morphology of LCs 
(see the working hypothesis proposed by Candia et al.\ 2003 to explain the 
bolometric behavior of SN~2000cx). The distribution of $^{56}$Ni or the 
degree of mixing are possibly other important variables.\\
The nature of these differences arising in the context of the ``explosion'' 
of a white dwarf at the Chandrasekar limit is still an open issue and requires 
more theoretical work.

\section{Conclusions}
\label{concl}
We have presented optical and NIR infrared photometry of the Type Ia SN~2000E
spanning a period from $\sim 16$ days ($\sim 7$ days in the NIR) before 
maximum $B$ light to $\sim 218$ days ($\sim 126$ days in the NIR) after it,
with a good coverage at peak. 
These data have been complemented with optical spectra from 6 days before 
maximum light to 122 days after it.
However, previous detections set the rise time to at least $ \ge 18$ days. 
The available optical data show slight discrepancies 
between different instruments up to $\sim 0.1$ mag. Although these differences 
have a negligible effect on our conclusions it still seems imperative to 
always acquire high accuracy photometry.
Our main conclusions have been:
\begin{enumerate}
\item SN~2000E is clearly reddened; we estimated an $E(B-V) \sim 0.5$ mag.
\item It is a slowly declining overluminous SN, with  $\Delta m_{15}(B)=0.94
 \pm 0.05$ and  $\Delta = -0.3,-0.5$ (as defined in the MLCS method).
\item Within the group of overluminous SNe, it is quite a typical 
representative, exhibiting
 a double-peak morphology at $RIJHK$, and a shoulder both at $V$ and in the
 bolometric light curve at the epoch of the secondary IR maximum.
\item SN~2000E is a spectroscopically ``normal'' SN Ia;
  the spectra of slow decliners with the same $\Delta m_{15}(B)$
   trace a sequence where the lines of SII and SiII display a decreasing 
   intensity down to the spectroscopically peculiar SN~1991T.
\item SN~2000E fits a phenomenological picture of SNe Ia in which more luminous
 objects exhibit  slowly-declining optical light curves and more strongly 
 developed two-peak morphology of NIR light curves,
 with imprints at even shorter wavelengths (down to $V$) and in the bolometric 
light curve.
\item The shoulder in the bolometric light curves at the epoch of the 
secondary IR peak and the width of wavelength range involved (from $V$ to $K$) 
clearly indicate that the double-peak morphology in the NIR light curves is 
not due to mere redistribution of flux from blue to red, but implies some sort 
of photospheric cooling and
 opacity variation with rapid escape of previously stored energy.
\item In a $J-H$ vs.\ $H-K$ diagram, all SNe Ia exhibit a continuously 
evolving  NIR excess with a similar crossing of the locus of main sequence 
during the period spanning the date of the main NIR peak and that of the 
minimum between peaks.
\item Based on the optical and NIR data, we revised the distance to the host 
galaxy,  NGC~6951, through the MLCS method, decline-rate estimated intrinsic 
brightness, and absolute NIR magnitudes 10 and 13.75 days past maximum $B$ 
light. We found an average value $\mu_{0} = 32.14 \pm 0.21$ mag, based on
the Sandage-Saha Cepheid calibration.
\item At the estimated distance, the peak bolometric luminosity is
 $\sim 1.998 \times 10^{43}$ ergs s$^{-1}$ corresponding to a mass of 
$^{56}$Ni
 of $0.9$ $M_{\odot}$ (assuming a rise time of 18 days).
\item If the photosphere can be approximated by a blackbody during the LC 
rising phase, then we find that it starts receding well before maximum light.
The approximation is probably no longer valid already one week before maximum 
$B$ light. 
\end{enumerate}

\begin{acknowledgements}

This paper is partially based on observations made with the Italian Telescopio
Nazionale Galileo (TNG) operated on the island of La Palma by the Centro
Galileo Galilei of the INAF (Istituto Nazionale di Astrofisica) 
at the Spanish Observatorio del Roque de los Muchachos of the Instituto de 
Astrofisica de Canarias. Most of the optical data are based  on observations 
made with the 0.7m (TNT) telescope operated on Teramo (Italy) by the 
INAF-Osservatorio Astronomico di Collurania-Teramo and with the 1.82m 
Copernico telescope operated on Asiago (Italy) by the INAF-Osservatorio 
Astronomico di Padova. The infrared data collected with the 
AZT-24 are products of the SWIRT, a joint project of the Astronomical 
Observatories of Collurania-Teramo (Italy), Pulkovo (Russia) and Rome (Italy).
We thank Adam Riess for kindly providing us with the updated templates of MLCS.
 We acknowledge support from the Italian Ministry for Education,
University and Research (MIUR) through grant Cofin 2000, under the scientific 
project ``Stellar Observables of Cosmological Relevance'', Cofin 
MM0290581 and Cofin 2002, under the scientific project ``Nubi di Magellano: 
test fondamentale per la teoria di evoluzione stellare e scala di distanza 
cosmica''.

\end{acknowledgements}



\begin{thebibliography}{}

\bibitem[Arnett(1982)]{arn82} Arnett, W.\ D.\ 1982, \apj, 253, 785 

\bibitem[Bessel (1979)]{bes79} Bessel, M.\ S.\ 1979, \pasp, 91, 589

\bibitem[Branch 1992]{bra92} Branch, D.\ 1992, \apj, 392, 35

\bibitem[]{} Burnstein, D., \& Heiles, C.\ 1982, \aj, 87, 1165

\bibitem[]{} Candia, P., et al.\ 2003, \pasp, in press

\bibitem[Cardelli et al.(1989)]{car89} Cardelli, J.\ A., Clayton, G.\ C., \& 
Mathis, J.\ S.\ 1989, \apj, 345, 245

\bibitem[Contardo(2000)]{con00} Contardo, G., Leibundgut, B., \& Vacca, W.\ 
D.\ 2000, \aap, 359, 876

\bibitem[]{} D'Alessio, F., et al.\ 2000, in Proc.\ SPIE 4008, ed.\ I.\ Masanori
 \& A.\ F.\ Moorwood (SPIE Press), 748

\bibitem[]{} Di Carlo, E., et al.\ 2002, \apj, 573, 144

\bibitem[]{} Elias, J.\ H., Frogel, J.\ A., Hackwell, J.\ A., \&
 Persson, S.\ E., 1981, \apjl, 251, 13

\bibitem[]{} Evans, R., \& Corso, G., 2000, IAUC 7359

\bibitem[]{} Freedman, W., et al.\ 2001, \apj, 553, 47

\bibitem[]{} Gibson, B.\ K., et al.\ 2000, \apj, 529, 723

\bibitem[]{} Hamuy, M., Phillips, M.\ M., Suntzeff, N.\ B., Schommer, 
R.\ A., Maza, J., \& Aviles, R.\ 1996a, \aj, 112, 2391

\bibitem[]{} Hamuy et al.\ 1996b, \aj, 112, 2408

\bibitem[Hamuy et al.\ 2002]{mario99ee} Hamuy M., et al.\ 2002, \aj, 124, 417

\bibitem[]{} Hamuy, M., et al.\ 2003, \aj, in Press 

\bibitem[]{} H\"{o}flich, P., Khokhlov, A.\ M., \& Wheeler, J.\ C.\ 1995, \apj,
444, 831

\bibitem[]{} Hunt, L.\ K., Mannucci, F., Testi, L., Migliorini, S., 
   Stanga, R.\ M., Baffa, C., Lisi, F. \& Vanzi, L.\ 1998, \aj, 115, 2594.

\bibitem[Kohno et al.(1999)]{koh99} Kohno, K., Kawabe, R., \& Vila-Vilar\'o, 
B.\ 1999, \apj, 511, 157

\bibitem[Koornneef 1983]{koo83} Koornneef, J.\ 1983, \aap, 128, 84

\bibitem[]{} Krisciunas, K., Hastings, N.\ C., Loomis, K., McMillan, R., Rest, 
A., Riess, A.\ G. \& Stubbs, C.\ 2000, \apj, 539, 658 

\bibitem[]{} Krisciunas, K., et al.\ 2001, \aj, 122, 1616

\bibitem[]{} Krisciunas, K., et al.\ 2003, \aj, 125, 166 

\bibitem[Landolt (1992)]{lan92} Landolt, A.\ U.\ 1992, \aj, 104, 340

\bibitem[Leibundgut et al.\ 1991]{bruno90n} Leibundgut, B., Kirshner, R.\ P., 
Filippenko, A.\ V., Shields, J.\ C., Foltz, C.\ B., Phillips, M.\ M., \& 
Sonneborn, G.\ 1991, \apjl,  371, 23

\bibitem[]{} Leibundgut, B., \& Pinto, P.\ A. 1992, \apj, 401,49

\bibitem[]{} Li, W., Filippenko, A. V., Treffers, R. R., Riess A. G., Hu, J., 
\&  Qiu, Y. L. 2001, \apj, 546, 734

\bibitem[]{} Lira, P. 1995, Master thesis, Univ.\ Chile 

\bibitem[]{} Lira, P., Suntzeff, N.\ B., Phillips, M.\ M., \& Hamuy, M.\ 1998, 
\aj, 115, 234

\bibitem[Lisi et al 1996]{lis96} Lisi, F., et al.\ 1996, \pasp, 108, 364

\bibitem[Mazzali 1995]{mazzali} Mazzali, P.\ A., Danziger, I.\ J. \& Turatto 
M.\ 1995, \aap, 297, 509

\bibitem[]{} Meikle, W.\ P.\ S., et al.\ 1996, \mnras, 281, 263

\bibitem[]{} Meikle, W.\ P.\ S.\ 2000, \mnras, 314, 782 

\bibitem[]{} Nugent, P., Kim, A., \& Perlmutter, S.\ 2002, \pasp, 114, 803

\bibitem[Patat et al.\ 1996]{nando} Patat, F., Benetti, S., Cappellaro, E., 
Danziger, I.\ J., della Valle, M., Mazzali, P.\ A. \& Turatto, M.\ 1996, \mnras, 278, 111

\bibitem[]{} Phillips, M.\ M.\ 1993, \apjl, 413, L105

\bibitem[]{} Phillips, M.\ M., Lira, P., Suntzeff, N.\ B., Schommer, R.\ A., 
Hamuy, M. \& Maza, J.\ 1999, \aj, 118, 1766 

\bibitem[]{} Phillips, M. M., et al.\ 2002, preprint (astro-ph/0211100)

\bibitem[]{} Phillips, M.\ M., et al.\ 2003, 
 in From Twilight to Highlight - The Physics of Supenovae, 
 ESO/MPA/MPE Workshop, in press

\bibitem[]{} Piersanti, L., Gagliardi, S., Iben, I.\ J., \& Tornamb\'{e}, A.\ 
 2003, \apj, 583, 885

\bibitem[]{} Pinto, P.\ A., \& Eastman, R.\ G.\ 2000a, \apj, 530, 744

\bibitem[]{} Pinto, P.\ A., \& Eastman, R.\ G.\ 2000b, \apj, 530, 757

\bibitem[]{} Riess, A.\ G., Press, W.\ H. \& Kirshner, R.\ P.\ 1996, \apj, 473, 88

\bibitem[]{} Riess, A.\ G., et al.\ 1998, \aj, 116, 1009

\bibitem[]{} Riess, A.\ G., et al.\ 1999, \aj, 118, 2675

\bibitem[]{} Saha, A., Sandage, A., Labhardt, L., Tammann, G.\ A., Macchetto, 
F.\ D. \& Panagia, N.\ 1997, \apj, 486, 1

\bibitem[]{} Salvo, M.\ E., Cappellaro, E., Mazzali, P.\ A., Benetti, S., 
Danziger, I.\ J., Patat, F.,\& Turatto, M.\ 2001, \mnras, 321, 254

\bibitem[]{} Sandage, A., Saha, A., Tammann, G.\ A., Labhart, L., Panagia, N., \& Macchetto, F.\ D.\ 1996, \apjl, 460, L15

\bibitem[]{} Schlegel, D., Finkbeiner, D., \& Davis, M.\ 1998, \apj, 500, 525 

\bibitem[]{} Stritzinger, M., et al.\ 2002, \aj, 124, 2100

\bibitem[]{} Strolger, L.\ G., et al.\ 2002, \aj, 124, 2905

\bibitem[]{} Suntzeff, N.\ B.\ 1996, in IAU
 Colloquium 145, Supernovae and Supernovae Remnants, ed.\ R. McCray \&
 Z.\ Wang (Cambridge: Cambridge  University Press), 41

\bibitem[]{} Suntzeff, N.\ B.\ 2003, in Proc.\ of the ESO/MPA/MPE Workshop, 
From Twilight to Highlight: The Physics of Supernovae, ed.\ B.\ Leibundgut \&
W.\ Hillenbrandt (Springer-Verlag), preprint (astro-ph/0212561)

\bibitem[]{} Turatto, M., Benetti, S., \& Cappellaro, E.\ 2003, 
in Proc.\ of the ESO/MPA/MPE Workshop, From Twilight to 
Highlight: The Physics of Supernovae, ed.\ B.\ Leibundgut \&
W.\ Hillenbrandt (Springer-Verlag), preprint 
(astro-ph/0211219)

\bibitem[]{} Turatto, M., Galletta, G., \& Cappellaro, E.\ 2000, IAUC 7351

\bibitem[]{} Valentini, G., et al.\ 2000, IAUC 7351.

\bibitem[Vinko et al.(2001)]{vin01} Vink\'o, J., et al.\ 
2001, \aap, 372, 824

\bibitem[Wheeler \& Benetti 2000]{wb} Wheeler, J.\ C., \& Benetti, S.\ 2000, 
$18^{th}$ Chapter of Allen's Astrophysical Quantities (IV$^{th}$ 
edition, AIP Press), 451

\end{thebibliography}
\end{document}